\pdfoutput=1
\documentclass[AMA,STIX1COL]{WileyNJD-v2}

\articletype{Article Type}%

\received{26 April 2016}
\revised{6 June 2016}
\accepted{6 June 2016}

\begin{document}

\title{Evaluating the Robustness of Targeted Maximum Likelihood Estimators via Realistic Simulations in Nutrition Intervention Trials}

\author[1]{Haodong Li}
\author[2]{Sonali Rosete}
\author[3]{Jeremy Coyle}
\author[4]{Rachael V. Phillips}
\author[5]{Nima S. Hejazi}
\author[6]{Ivana Malenica}
\author[7]{Benjamin F. Arnold}
\author[8]{Jade Benjamin-Chung}
\author[9]{Andrew Mertens}
\author[10]{John M. Colford Jr}
\author[11]{Mark J. van der Laan}
\author[12]{Alan E. Hubbard*}

\authormark{Li \text{et al}}

\address[1,2,3,4,5,6,9,10,11,12]{\orgdiv{Divisions of Epidemiology \& Biostatistics}, \orgname{University of California, Berkeley}, \orgaddress{2121 Berkeley Way Rm 5302, \city{Berkeley}, \state{California}, \country{USA}, 94720-7360}}

\address[7]{\orgdiv{Proctor Foundation}, \orgname{University of California, San Francisco}, \orgaddress{490 Illinois Street, \city{San Francisco}, \state{California}, \country{USA}, 94158}}

\address[8]{\orgdiv{Epidemiology \& Population Health}, \orgname{Stanford University}, \orgaddress{150 Governor's Lane, \city{Stanford}, \state{California}, \country{USA}, 94305-5101}}

\corres{*Alan E Hubbard, Division of Biostatistics, University of California, Berkeley, 2121 Berkeley Way Rm 5302, Berkeley, CA 94720-7360 \email{hubbard@berkeley.edu}}

\abstract[Summary]{Several recently developed methods have the potential to harness machine learning in the pursuit of target quantities inspired by causal inference, including inverse weighting, doubly robust estimating equations and substitution estimators like targeted maximum likelihood estimation.  There are even more recent augmentations of these procedures that can increase robustness, by adding a layer of cross-validation (cross-validated targeted maximum likelihood estimation and double machine learning, as applied to substitution and estimating equation approaches, respectively). While these methods have been evaluated individually on simulated and experimental data sets, a comprehensive analysis of their performance across ``real-world'' simulations have yet to be conducted. 
In this work, we benchmark multiple widely used methods for estimation of the average treatment effect using ten different nutrition intervention studies data. A realistic set of simulations, based on a novel method, highly adaptive lasso, for estimating the data-generating distribution that guarantees a certain level of complexity (undersmoothing) is used to better mimic the complexity of the true data-generating distribution.   We have applied this novel method for estimating the data-generating distribution by individual study and to subsequently use these fits to simulate data and estimate treatment effects parameters as well as their standard errors and resulting confidence intervals. Based on the analytic results, a general recommendation is put forth for use of the cross-validated variants of both substitution and estimating equation estimators. We conclude that the additional layer of cross-validation helps in avoiding unintentional over-fitting of nuisance parameter functionals and leads to more robust inferences.
}


\maketitle

\section{Introduction}\label{intro}

Epidemiological studies, particularly based on randomized trials, often aim to estimate the average treatment effect (ATE), or another causal parameter of interest, to understand the effect of a health intervention or exposure on an outcome of interest. Most commonly, in observational studies, inverse probability of treatment weighted (IPTW) estimation and its variants have been used for this purpose ~\cite{HorvitzGSRF1952, ipw_hajek_1971, AronowEACED2017}. Alternative estimators for causal inference include substitution (or direct) estimators based on G-computation~\cite{RobinsNACIJ1986,YuCCGFD2002, DanielGECED2011, WangGATEJ2017}, those based on the approach of estimating equations (EE)   \cite{RobinsMSMS2000,LaanUMCL2003}, including IPTW and its augmented variant (A-IPTW), and substitution estimators developed within the framework of targeted learning (TL) (we also refer to targeted maximum likelihood estimator, TMLE, a product of this framework \cite{LaanTLCI2011}).  The latter of these has seen increasing use in recent years, both in biostatistical methodological research and applied public health and medical research \cite{PetersenAIUTJ2017,SkeemCPSOS2017,RoseRMLVO2018,PlattTERCJ2018,NeafseyGDPEN2015}. In Table~\ref{tab:lit_summary}, we provide a list of studies that have examined the relative performance of TL-based and competing estimators (mainly against EE-based methods), including a summary of whether the results suggested superior, neutral, or poorer relative performance of TL-based estimators in comparison to other estimators (the ``Pro/Con'' column). Thus, while this work is contextualized within dozens of previous studies, few such studies performed ``realistic'' simulations, and even fewer compared several variants of TL estimators alongside corresponding EE approaches. For example, in Zivich and Breskin's paper \cite{ZivichMLCIM2021}, the authors compared G-computation, IPTW, A-IPTW, TMLE and double cross-fit estimators with data generated from predefined parametric models. Exceptions are efforts that used the proposed realistic bootstrap~\cite{PetersenDRVPF2012} to evaluate the performance for data-generating distributions modeled semiparametrically (using ensemble machine learning) from an existing data set. These include a study of estimating variable importance under positivity violations using collaborative targeted maximum likelihood estimation (C-TMLE)~\cite{PirracchioCTMLJ2018}. In this paper, we use an augmentation of this proposed methodology to examine the relative performance of several versions of both TL and EE estimators in ten realistic data simulations, each based on data collected as part of the Knowledge Integration (KI) database from the Bill \& Melinda Gates Foundation ~\cite{MertensCCCGJ2020}. In so doing, we provide a realistic survey, across both different data-generating distributions and different study designs, of the relative performance of estimators of causal parameters.

\begingroup\fontsize{9}{11}\selectfont

\begin{longtable}[t]{p{2cm}p{4cm}cp{7cm}c}
\caption{\label{tab:lit_summary}Overview of literature on comparison of TMLE and other estimators.  The Pro/Con column refers to a simple binary classification of the relative performance of the TMLE estimators reported in the paper, "Pro" indicating that the TMLE performed superior to other competing estimators.} \\
\toprule
Authors & Title & Year & Description of Reuslts & Pro/Con\\
\midrule
\cellcolor{gray!6}{Chatton, et al.\cite{ChattonGPSMJ2020}} & \cellcolor{gray!6}{G-computation, propensity score-based methods, and targeted maximum likelihood estimator for causal inference with different covariates sets: a comparative simulation study} & \cellcolor{gray!6}{2020} & \cellcolor{gray!6}{Article compares different semi-parametic approaches, including TMLE and matching, but finds G-computation performs relatively best.  Given their simulation, this was predictable because they simulated from a parametric model and used the same model for estimating the regression, thus showing the superiority of maximum likelihood estimation in parametric models. This is not a realistic setting.} & \cellcolor{gray!6}{Con}\\
Talbot and Beaudoin \cite{TalbotGDRBM2020} & A generalized double robust Bayesian model averaging approach to causal effect estimation with application to the Study of Osteoporotic Fractures & 2020 & Proposed a Generalized Bayesian Causal Effect Estimation (GBCEE), which outperformed double robust alternatives(including C-TMLE). Also showed ``target'' A-IPTW is superior than C-TMLE in a non-realistic setting(only using true confounders). & Con\\
\cellcolor{gray!6}{Zivich and Breskin \cite{ZivichMLCIM2021}} & \cellcolor{gray!6}{Machine learning for causal inference: on the use of cross-fit estimators} & \cellcolor{gray!6}{2020} & \cellcolor{gray!6}{A simulation study assessing the performance of G-computation, IPW, AIPW, TMLE, doubly robust cross-fit (DC) AIPW and DC-TMLE. With correctly specified parametric models, all of the estimators performed well. When used with machine learning, the DC estimators outperformed other estimators.} & \cellcolor{gray!6}{Neutral}\\
Ju, et al. \cite{JuSCTLF2019} & Scalable collaborative targeted learning for high-dimensional data & 2019 & Results from simulations suggested superior performance of C-TMLE relative to both A-IPTW and non-collaborative ("standard") TMLE estimators. & Pro\\
\cellcolor{gray!6}{Ju, et al. \cite{JuAPSTJ2019}} & \cellcolor{gray!6}{On adaptive propensity score truncation in causal inference} & \cellcolor{gray!6}{2019} & \cellcolor{gray!6}{By adaptively truncating the estimated propensity score with a more targeted objective function, the Positivity-C-TMLE estimator achieves the best performance for both point estimation and confidence interval coverage among all estimators considered.} & \cellcolor{gray!6}{Pro}\\
\addlinespace
Bahamyirou, et al. \cite{BahamyirouUDPBJ2019} & Understanding and diagnosing the potential for bias when using machine learning methods with doubly robust causal estimators & 2019 & Simulation results  showed superior performance of C-TMLE and TMLE relative to IPTW. & Pro\\
\cellcolor{gray!6}{Wei, et al. \cite{WeiDTLAS2019}} & \cellcolor{gray!6}{A Data-Adaptive Targeted Learning Approach of Evaluating Viscoelastic Assay Driven Trauma Treatment Protocols} & \cellcolor{gray!6}{2019} & \cellcolor{gray!6}{C-TMLE outperformed the other doubly robust estimators (IPTW, A-IPTW, stabilized IPTW, TMLE) in the simulation study.} & \cellcolor{gray!6}{Pro}\\
Rudolph, et al. \cite{RudolphCSDED2019} & Complier Stochastic Direct Effects: Identification and Robust Estimation & 2019 & Showed that the EE and TMLE estimators have advantages over the IPTW estimator in terms of efficiency and reduced reliance on correct parametric model specification. & Pro\\
\cellcolor{gray!6}{Pirracchio, et al. \cite{PirracchioCTMLJ2018}} & \cellcolor{gray!6}{Collaborative targeted maximum likelihood estimation for variable importance measure: Illustration for functional outcome prediction in mild traumatic brain injuries} & \cellcolor{gray!6}{2018} & \cellcolor{gray!6}{Showed much more robust performance of C-TMLE relative to TMLE using the same type of realistic parametric bootstrap as used in this paper.  This was under severe near-positivity violations.} & \cellcolor{gray!6}{Pro}\\
Luque‐Fernandez, et al. \cite{Luque-FernandezTMLE2018} & Targeted maximum likelihood estimation for a binary treatment: A tutorial & 2018 & Showed relatively superior performance of TMLE when compared with A-IPTW estimator in terms of bias. & Pro\\
\addlinespace
\cellcolor{gray!6}{Levy, et al. \cite{LevyFMTEJ2021}} & \cellcolor{gray!6}{A Fundamental Measure of Treatment Effect Heterogeneity} & \cellcolor{gray!6}{2018} & \cellcolor{gray!6}{Showed the advantage of CV-TMLE over TMLE in that TMLE was affected by overfitting while CV-TMLE appeared unaffected.} & \cellcolor{gray!6}{Pro}\\
Schuler and Rose \cite{SchulerTMLEJ2017}& Targeted maximum likelihood estimation for causal inference in observational studies & 2017 & Showed  superior performance of TMLE relative to misspecified  parametric models. & Pro\\
\cellcolor{gray!6}{Pang, et al. \cite{PangEEPSN2016}} & \cellcolor{gray!6}{Effect Estimation in Point-Exposure Studies with Binary Outcomes and High-Dimensional Covariate Data--A Comparison of Targeted Maximum Likelihood Estimation and Inverse Probability of Treatment Weighting} & \cellcolor{gray!6}{2016} & \cellcolor{gray!6}{Showed relatively superior performance for the TMLE  to IPTW, which showed greater instability when positivity violations occurred.} & \cellcolor{gray!6}{Pro}\\
Schnitzer, et al. \cite{SchnitzerVSCCM2016} & Variable Selection for Confounder Control, Flexible Modeling and Collaborative Targeted Minimum Loss-Based Estimation in Causal Inference & 2016 & Using IPTW with flexible prediction for the propensity score can result in inferior estimation, while TMLE and C-TMLE may benefit from flexible prediction and remain robust to the presence of variables that are highly correlated with treatment. & Pro\\
\cellcolor{gray!6}{Zheng, et al. \cite{ZhengDREEM2016}} & \cellcolor{gray!6}{Doubly Robust and Efficient Estimation of Marginal Structural Models for the Hazard Function} & \cellcolor{gray!6}{2016} & \cellcolor{gray!6}{Showed that the TMLE for marginal structual model (MSM) for a hazard function has relatively superior performance. The bias reduction over a misspecified IPTW or Gcomp estimator is clear in the simulation studies even for a moderate sample size.} & \cellcolor{gray!6}{Pro}\\
\addlinespace
Schnitzer, et al. \cite{SchnitzerDREEJ2016} & Double robust and efficient estimation of a prognostic model for events in the presence of dependent censoring & 2016 & This study demonstrated that even when the analyst is ignorant of the true data generating form, TMLE with Super Learner can perform about as well as IPTW or TMLE with correct parametric model specification. & Pro\\
\cellcolor{gray!6}{Kreif, et al. \cite{KreifETEMO2016}} & \cellcolor{gray!6}{Evaluating treatment effectiveness under model misspecification: A comparison of targeted maximum likelihood estimation with bias-corrected matching} & \cellcolor{gray!6}{2014} & \cellcolor{gray!6}{Examined the relative performance of TMLE, EE and  matching estimators showing superior performance of TMLE when the outcome regression is misspecified.} & \cellcolor{gray!6}{Pro}\\
Schnitzer, et al. \cite{SchnitzerEBGIJ2014} & Effect of breastfeeding on gastrointestinal infection in infants: A targeted maximum likelihood approach for clustered longitudinal data & 2014 & Compared TMLE with IPTW and G-computation, under the plausible scenario of being given transformed versions of the confounders. Only TMLE with Super Learner was able to unbiasedly estimate the parameter of interest. & Pro\\
\cellcolor{gray!6}{Gruber and van der Laan \cite{GruberATMLM2013}} & \cellcolor{gray!6}{An Application of Targeted Maximum Likelihood Estimation to the Meta-Analysis of Safety Data} & \cellcolor{gray!6}{2013} & \cellcolor{gray!6}{Reported superiority of both TMLE and A-IPTW to misspecified parametric models, but the data-generating distributions used resulted in little difference between the semi-parametric approaches.} & \cellcolor{gray!6}{Neutral}\\
Lendle, et al. \cite{LendleTMLEA2013} & Targeted maximum likelihood estimation in safety analysis & 2013 & Showed superior performance of TMLE and C-TMLE relative to A-IPTW estimators in the context of  positivity violations. & Pro\\
\addlinespace
\cellcolor{gray!6}{Díaz and van der Laan \cite{DiazTDAED2013}} & \cellcolor{gray!6}{Targeted Data Adaptive Estimation of the Causal Dose Response Curve} & \cellcolor{gray!6}{2013} & \cellcolor{gray!6}{Showed relatively superior performance of CV-TMLE relative to CV-A-IPTW estimators, especially in the presence of empirical violations of the positivity assumption.} & \cellcolor{gray!6}{Pro}\\
Schnitzer, et al. \cite{SchnitzerTMLEJ2013} & Targeted maximum likelihood estimation for marginal time-dependent treatment effects under density misspecification & 2013 & In the simulation study, TMLE did not produce a reduction in finite-sample bias or variance for correctly specified densities compared with the G-computation estimator, but it had much better performance than G-computation when the outcome model was misspecified. & Neutral\\
\cellcolor{gray!6}{Petersen, et al. \cite{PetersenDRVPF2012}} & \cellcolor{gray!6}{Diagnosing and responding to violations in the positivity assumption} & \cellcolor{gray!6}{2012} & \cellcolor{gray!6}{Showed  superior performance of TMLE relative to misspecified  parametric models, in comparison with A-IPTW, IPTW and G-computation.} & \cellcolor{gray!6}{Pro}\\
van der Laan and Gruber \cite{vanderLaanTMLB2012} & Targeted Minimum Loss Based Estimation of Causal Effects of Multiple Time Point Interventions & 2012 & In the setting of multiple time point interventions, showed TMLE outperformed IPTW and MLE estimators. & Pro\\
\cellcolor{gray!6}{Porter, et al. \cite{PorterRPTM2011}} & \cellcolor{gray!6}{The relative performance of targeted maximum likelihood estimators} & \cellcolor{gray!6}{2011} & \cellcolor{gray!6}{Showed relatively superior performance of C-TMLE relative to A-IPTW estimators particularly when there are covariates that are strongly associated with the missingness, while while being weakly or not at all associated with the outcome.} & \cellcolor{gray!6}{Pro}\\
\addlinespace
Wang, et al. \cite{WangFQTLJ2011} & Finding Quantitative Trait Loci Genes with Collaborative Targeted Maximum Likelihood Learning & 2011 & Based on actual genetic data, results suggested greater robustness of findings using C-TMLE relative to parametric approaches for high throughput genetic data. & Pro\\
\cellcolor{gray!6}{Díaz and van der Laan \cite{MunozPICEJ2012}} & \cellcolor{gray!6}{Population Intervention Causal Effects Based on Stochastic Interventions} & \cellcolor{gray!6}{2011} & \cellcolor{gray!6}{Paper focused on new estimators for stochastic (e.g., shift) interventions relevant to estimating causal effects of continuous interventions.   In their simulation, they did not observe significant differences between the TMLE and the A‐IPTW.} & \cellcolor{gray!6}{Neutral}\\
Gruber and van der Laan \cite{GruberACTM2010} & An application of collaborative targeted maximum likelihood estimation in causal inference and genomics & 2010 & Showed more robust performance in high-dimensional simulations comparing TMLE to estimating equation approaches (A-IPTW). & Pro\\
\cellcolor{gray!6}{Stitelman and van der Laan \cite{StitelmanCTML2010}} & \cellcolor{gray!6}{Collaborative Targeted Maximum Likelihood for Time to Event Data} & \cellcolor{gray!6}{2010} & \cellcolor{gray!6}{The results show that, compared with TMLE, IPTW and A-IPTW, the C-TMLE method does at least as well as the best estimator under every scenario and, in many of the more realistic scenarios, behaves much better than the next best estimator in terms of both bias and variance.} & \cellcolor{gray!6}{Pro}\\
Moore and van der Laan \cite{MooreCARTJ2009} & Covariate adjustment in randomized trials with binary outcomes: targeted maximum likelihood estimation & 2009 & Demonstrated how the use of covariate information in randomized clinical trials could use the TMLE framework, which results in improved performance, without bias, relative to standard methods. & Pro\\
\addlinespace
\cellcolor{gray!6}{Rose and van der Laan \cite{RoseSOWCS2008}} & \cellcolor{gray!6}{Simple Optimal Weighting of Cases and Controls in Case-Control Studies} & \cellcolor{gray!6}{2008} & \cellcolor{gray!6}{IPTW method for causal parameter estimation was outperformed in conditions similar to a practical setting by the new case-control weighted TMLE methodology.} & \cellcolor{gray!6}{Pro}\\
\bottomrule
\end{longtable}
\endgroup{}

\section{Background}\label{background}
As large and complex data sets have become increasingly more commonplace, the habitual use of parametric approaches is encountering more data science research for which they are ill-suited. .  This has led to machine learning (ML) taking a more central role in deriving estimators of causal impacts in very big statistical models (semiparametric).  The theory for the use of ML in the estimators discussed herein has been continuously refined, from developing double robust estimators (both A-IPTW and TMLE substitution estimators) to augmentations of these estimators that are more robust to the overfitting potentially introduced by flexible ML fits.  The latter modifications to the original estimators are the cross-validated TMLE (or CV-TMLE, chapter 27 in van der Laan\cite{LaanTLCI2011} and Zheng \cite{ZhengATCTN2010}), and subsequently the proposal for an analogous modification to estimating equation approaches (double machine learning or cross-fitting \cite{ChernozhukovDDMLF2018}).   

While simulation studies have investigated all of these estimators, they have yet to be analyzed together in a single series of realistic simulation studies. Here, we seek to determine how well these estimators perform in realistic settings, under which conditions they perform best, which augmentations provide the most robustness, and whether or not the results support more general recommendations. In addition, there exist other choices of target parameter when the one being analyzed fails to have adequate performance for any of the competing estimators, such as realistic rules \cite{BembomPIIR2007}.  A recently developed  machine learning algorithm (the highly adaptive lasso; HAL \cite{BenkeserHALE2016}), is potentially an important improvement in estimating realistic DGDs for simulation studies such as ours. It can be optimally undersmoothed to dependably generate realistic estimates of the actual data generating distributions. HAL is particularly well suited to these types of simulations, as it uses a very large nonparametric model and can be tuned to be as flexible as the data support.  In this paper, we explore the use of HAL as a basis in conducting realistic data-inspired simulations.  The results suggest the proposed use of HAL for realistic data-generating simulations could provide a general method for choosing between machine-learning-based estimators for a particular parameter and data set.

We first introduce the data sets that were selected to motivate our realistic simulations, describe the steps taken for simulating data, including a short description of the estimators tested, and discuss the results. The simulations suggest a general recommendation for the use of an additional layer of cross-validation (CV-TMLE and double machine learning) to ensure robust inference in finite samples.

\section{Methods}\label{method}

\subsection{Study Selection} 
We utilized data from ten nutrition intervention trials conducted in Africa and South Asia. In all studies, the measured outcome was a height-to-age Z-score for children from birth to 24 months, which was calculated using World Health Organization (WHO) 2006 child growth standards~\cite{WHOGrowth2006}. Details about the resulting composite data, study design and data processing, can be found in companion technical reports \cite{MertensCCCGJ2020,MertensCWCSJ2020,Benjamin-ChungECLGJ2020}. All interventions were nutrition-based, and for the purposes of this analysis, multi-level interventions were simplified to a binary treatment variable (e.g. nutrition intervention - yes/no). Although different baseline covariates were measured among these studies, there was significant overlap. The sample size of each study is shown in Table \ref{tab_study}. We anonymized the study IDs and removed the location information due to confidentiality concerns. Details on each study can be found in the shuffled list in Section B of the Appendix.

\begin{table}

\caption{\label{tab_study}Dimensions of datasets of Nutrition Intervention Trials, with $n$ representing the number of children in sample and $p$ being the number of covariates.}
\centering
\begin{tabular}[t]{>{\centering\arraybackslash}p{2cm}>{\centering\arraybackslash}p{3cm}>{\centering\arraybackslash}p{3cm}}
\toprule
Study ID & n & p\\
\midrule
1 & 418 & 20\\
2 & 4863 & 26\\
3 & 7399 & 22\\
4 & 1204 & 36\\
5 & 2396 & 42\\
\addlinespace
6 & 3265 & 18\\
7 & 1931 & 38\\
8 & 840 & 30\\
9 & 27275 & 42\\
10 & 5443 & 35\\
\bottomrule
\end{tabular}
\end{table}

\subsection{Data Processing}
Data from each study was cleaned and processed for this analysis. Our goal for defining the analysis data used to simulate is different from the goals of the original studies and thus our data processing might differ from that used in the resulting publications of the study results.  We note that the data are used to motivate the simulations, but, since we define the true DGD to be one that we estimate for each study, and at that point differences with the original study become irrelevant to our comparisons of estimators. Data was filtered down to the last height-to-age Z-score measurement taken at the end of each study for each subject. Subjects were dropped if either their treatment assignment ($A$) or outcome measurement ($Y$) were missing. For covariates ($W$) that were missing, those that were continuous and discrete were imputed using the median and mode, respectively. In both cases, missingness indicator variables were added to the data set for each covariate with missing rows. As mentioned above, the treatment assignment variable ($A$) was binarized if it consisted of more than two treatment arms. The control and treatment groups were originally assigned in each study as described in Section B of the Appendix.

\subsection{Simulation with Undersmoothed Highly Adaptive Lasso}
\subsubsection{Undersmoothed Highly Adaptive Lasso}
We adopted the undersmoothed highly adaptive lasso (HAL) method to generate data for simulations in a (nearly) nonparametric model, with as few assumptions as possible. HAL is a nonparametric regression estimator, which neither relies upon local smoothness assumptions nor is constructed using local smoothing techniques \cite{BenkeserHALE2016}. HAL has been shown to have competitive finite-sample performance relative to many other popular machine learning algorithms.  
With the assumption that the target parameter $\psi$ falls in the Donsker class of all cadlag functions (right-hand continuous, with left-hand limits) with finite variation norm, we have the following representation\cite{GillIEBS1995}:
$$
\psi(\bx) = \psi(0) + \sum_{s\subset \{1,2,...,p\}} \int_{0_s}^{\bx_s}\psi_s(du)
$$
where $\bx \in \mathbb{R}^p$ and $s$ denotes the indices of sections of $\psi$. 
Then let us further denote $\bx_s = (x_k: k \in s)$ as the subvector with support of $s$, and $\tilde{\bx}_{s,i}$ as the values of the subvector for the $i$\textsuperscript{th} observation. Now $\psi$ can be approximated by $\psi_m$ such that\cite{BenkeserHALE2016}:
$$
\psi_m(\bx) =  \psi(0)  + \sum_{s\subset \{1,2,...,p\}}\sum_{i=1}^{n} I(\tilde{\bx}_{s,i} \leq \bx_s)d\psi_{m,s,i}
$$

Now if we consider a model with the basis functions $\phi_{s,i} = I(\tilde{x}_{i,s} \leq x_s)$ as predictors and $d\psi_{m,s,i}$ as coefficients, we have\cite{BenkeserHALE2016}:
$$
\psi_{\beta} = \beta_0 + \sum_{s\subset \{1,2,...,p\}}\sum_{i=1}^{n} \beta_{s,i} \phi_{s,i}
$$

By the assumption of finite sectional variation norm (an entropy assumption required of all but two of the estimators), cross-validated TMLE and double-robust EE) we have the corresponding subspace $ \Psi_{n,M} = \{\psi_{\beta}: \beta_0 + \sum_{s\subset \{1,2,...,p\}}\sum_{i=1}^{n} \beta_{s,i} < M \}$\cite{BenkeserHALE2016}.

The HAL estimator starts with a very large number (at most $n*(2^p-1)$) of basis functions that are indicator functions with support at the observed data values. In practice, when some covariates are categorical or binary, the number of unique basis functions will be much fewer than the upper bound. Moreover,
to avoid overfitting, one can define a subspace of the linear model such that: $\sum_{s\subset \{1,2,...,p\}}\sum_{i=1}^{n} |\beta_{s,i}| \le C$, for submodels where the $L_1$-norm is bounded by $C$. The dimension of basis functions can be restricted. For example, one can consider only main-term indicators for each of the original predictors as well as all second order tensor products (interaction terms involving the main effect terms). One can use cross-validated selection of $C$ to optimize the fit of the model to future observations from the data-generating distribution (DGD). 

It has been recently shown that undersmoothing HAL (using a $C$ that is larger than that selected by cross-validation) can yield an asymptotically efficient estimators for functionals of the relevant portions of the DGD while preserving the same rate of convergence, and also solving the efficient score equation for any desired path-wise differentiable target feature of the data distribution~\cite{vanderLaanEEPDA2019}. As a consequence, an undersmoothed HAL results in an efficient plug-in estimator of the desired estimand, and moreover, it will also be efficient for any other smooth estimands of the data distribution~\cite{vanderLaanEEPDA2019}. This could serve as the basis for using HAL in our settings; that is, to estimate the DGD by HAL in a way that optimally preserves the relevant functionals. More intuitively, HAL, with the properly chosen $C$, will result in a DGD for simulations that is as close as one can get nonparametrically to the true DGD, without blowing up the variance of estimation. So, it creates a comparison that is as faithful as possible to the DGD of interest, itself represented by a single data set (experiment).  Thus, we argue that it can serve as the basis of a simulation where one wishes to compare estimators for the data in hand. We provide more rigorous justification below.

In our study, we only use the undersmoothed HAL to generate data without pre-specifying any parameter of interest. The stopping criterion for this undersmoothing process is to increase the initial bound $C_{cv}$ until the score equations formed by the product of basis functions and residuals are solved at the rate of $\frac{\sigma_n}{\sqrt{n} \log(n)}$ \cite{Laan_underHAL_2019}. Namely, for all ``non-trivial directions'' (combinations of $s,i$ with non-zero coefficients selected by the initial fit) we need:
\begin{align}
    |P_n \big(\phi_{s,i}(Y-\bar{Q}_{n,C})\big)| \leq \frac{\sigma_n}{\sqrt{n} \log(n)}
\end{align}
where $P_n$ is the empirical average function and $\sigma_n^2 = Var\big(\phi_{s,i}(Y-\bar{Q}_{n,C_{cv}})\big)$. Following the convention of notation in \cite{LaanTLCI2011}, we define $\bar{Q}_0 = \EE_{\PP_0}(Y|A,W)$ and $\bar{Q}_n$ as its estimate. Also, we use $Q_{W,0} = \PP_0(W), Q_0 = (\bar{Q}_0, Q_{W,0})$, and $Q$ denoting the possible value of true $Q_0$.\\
To justify this criterion, first fix $(s,i)$ and consider the target parameter $\Psi_{s,i}(Q_0) = P_0(\phi_{s,i}\bar{Q}_0) = \EE_{\PP_0}(\phi_{s,i}\bar{Q}_0) = 
\EE_{\PP_0}(\phi_{s,i}Y)$. (Notice that for this target parameter, we can treat $A$ as a member of $W$ so that $Q_0$ actually contains $\bar{Q}_0$ and $Q_{W',0} = \PP_0(A,W)$).
The last equality is true since $\EE_{\PP_0}(\phi_{s,i}Y) = \EE_{\PP_0}[\EE_{\PP_0}(\phi_{s,i}Y | A,W)] =
\EE_{\PP_0}[\phi_{s,i}\EE_{\PP_0}(Y|A,W)]$.
We claim that $\phi_{s,i}(Y - \bar{Q}_0)$ is a component of the efficient influence curve (EIC) of $\Psi_{s,i}({Q}_0)$, where we denote the EIC as $D^*_{s,i}({Q}_0)$. To prove this, we can start with the empirical estimator $\frac{1}{n}\sum_{k=1}^{n}\phi_{s,i}y_k$. Observe that
\begin{align*}
    \frac{1}{n}\sum_{k=1}^{n}\phi_{s,i}y_k - \Psi_{s,i}({Q}_0) = 
    \frac{1}{n}\sum_{k=1}^{n}(\phi_{s,i}y_k - \Psi_{s,i}({Q}_0))
\end{align*}
Thus the influence curve of the empirical estimator is $\phi_{s,i}Y - \Psi_{s,i}({Q}_0)$, denote it as $D^0_{s,i}({Q}_0)$. With it, we can obtain $D^*_{s,i}({Q}_0)$ by projecting $D^0_{s,i}({Q}_0)$ onto the tangent space $T(\PP_0)$~\cite{LaanTLCI2011}.
In addition, since $\PP(O) = \PP(Y, A, W) = \PP(Y|A,W)\PP(A,W)$, the tangent space $T(\PP_0)$ can be decomposed as: $T(\PP_0) = T_{Q}(\PP_0) = T_{Q_Y}(\PP_0) \oplus T_{Q_{A,W}}(\PP_0)$~\cite{LaanTLCI2011}. So the projection of $D^0_{s,i}({Q}_0)$ on $T(\PP_0)$ is equal to the sum of the projections of $D^0_{s,i}({Q}_0)$ on $T_{Q_Y}(\PP_0)$ and $T_{Q_{A,W}}(\PP_0)$, namely, $\sqcap(D^0_{s,i}({Q}_0) | T(\PP_0)) = \sqcap(D^0_{s,i}({Q}_0) | T_{Q_Y}(\PP_0)) + \sqcap(D^0_{s,i}({Q}_0) | T_{Q_{A,W}}(\PP_0))$. Thereby $D^*_{s,i}({Q}_0) = \sqcap(D^0_{s,i}({Q}_0) | T_{Q_Y}(\PP_0)) + \sqcap(D^0_{s,i}({Q}_0) | T_{Q_{A,W}}(\PP_0))$. Then
\begin{align*}
    \sqcap(D^0_{s,i}({Q}_0) | T_{Q_Y}(\PP_0))  
    =& D^0_{s,i}({Q}_0) - \EE_{\PP_0}(D^0_{s,i}({Q}_0)|A,W)~\cite{LaanTLCI2011}\\
    =& \phi_{s,i}Y - \Psi_{s,i}({Q}_0) - \EE_{\PP_0}(\phi_{s,i}Y - 
       \Psi_{s,i}({Q}_0)|A,W)\\
    =& \phi_{s,i}Y - \Psi_{s,i}({Q}_0) - \EE_{\PP_0}(\phi_{s,i}Y|A,W) +
       \Psi_{s,i}({Q}_0)\\
    =& \phi_{s,i}(Y - \EE_{\PP_0}(Y|A,W)) \\
    =& \phi_{s,i}(Y - \bar{Q}_0)\\
\mbox{and}\\
    \sqcap(D^0_{s,i}({Q}_0) | T_{Q_{A,W}}(\PP_0)) =& \EE_{\PP_0}(D^0_{s,i}({Q}_0)|A,W) ~\cite{LaanTLCI2011}\\
    =&  \EE_{\PP_0}(\phi_{s,i}Y - \Psi_{s,i}({Q}_0)|A,W) \\
    =& \EE_{\PP_0}(\phi_{s,i}Y |A,W) - 
       \Psi_{s,i}({Q}_0)
\end{align*}
So, $D^*_{s,i}({Q}_0) = \phi_{s,i}(Y - \bar{Q}_0) + \EE_{\PP_0}(\phi_{s,i}Y |A,W) - 
       \Psi_{s,i}({Q}_0)$.\\
Now we have proved that $\phi_{s,i}(Y - \bar{Q}_0)$ is a component of $D^*_{s,i}({Q}_0)$. Another observation from the calculation above is that $D^0_{s,i}({Q}_0) = D^*_{s,i}({Q}_0)$.\\
For different pairs of $(s,i)$, each $D^*_{s,i}(Q_{n,C})$ corresponds with an EIC for a particular target parameter $\Psi_{s,i}(Q_0)$.
For each plug-in estimator $\Psi_{s,i}({Q}_{n,C})$ being asymptotically linear we want at minimal $P_n D^*_{s,i}(Q_{n,C})=o_P(n^{-\frac{1}{2}})$ for every $(s,i)$, which is guaranteed by our choice of criterion. By doing so, we will also be solving $P_n (\sum_{s,i}\alpha(s,i)\phi_{s,i})(Y-\bar{Q}_{n,C})$ for any $\alpha$ vector with finite $L_1$-norm, which enables us to approximate any function of $(A,W)$ with rate approximately equal to $n^{-\frac{1}{3}}$~\cite{vanderLaanEEPDA2019}. So in this way we are rich enough to guarantee to solve any EIC that can be written as $f(A,W)(Y-\bar{Q}_0)$, thereby cover all EIC of features of Q. So the undersmoothing process essentially yields an estimator that is efficient for any target feature of Q that is pathwise differentiable. Combined with the fact that when the bias of an estimator is smaller than $\frac{se}{\max(10, \log{n})}$ then it has minimal impact on coverage, we choose (1) as the stopping criterion based on the proof above.

The specific procedure is stated as follows:

Step 1. Fit the HAL with $L_1$-norm, obtain the set of basis functions and a starting value of $\lambda$, denote it as $\lambda_{cv}$. This $\lambda_{cv}$ is a CV-optimal value of the penalty parameter returned by the hal9001 package \cite{coyle2021hal9001-rpkg, HejaziHSHAS2020}, which is essentially from the ``cv.glmnet'' function with 10-fold cross-validation.

Step 2. Calculate the absolute value of the normalized score equations for each direction:
$$
\big|\frac{\frac{1}{n}\sum_{k=1}^{n}[(y_{k} - \hat{y}_{\lambda})\phi_{s,i}]}
{\sqrt{\frac{1}{n}\sum_{k=1}^{n}[(y_{k} - \hat{y}_{\lambda_{cv}})\phi_{s,i} - 
\frac{1}{n}\sum_{k=1}^{n}(y_{k} - \hat{y}_{\lambda_{cv}})\phi_{s,i}]^2}}\big|
$$

Step 3. Take the maximum of this value from all subsets. Compare the max with $\frac{1}{\sqrt{n} \log(n)}$. If the max is larger, then increase the bound $C$ (i.e. decrease the value of the penalty parameter $\lambda$) and refit the HAL.

Step 4. Repeat 1,2,3 until the stopping criterion is satisfied.

In addition, we speed up the algorithm by controlling the number of basis functions in the initial HAL fits. First, we set the maximum interaction degree to $\mathbb{I}(p \geq 20)*2 +  \mathbb{I}(p < 20)*3$, where $p$ is the number of covariates. Second, we use binning method to restrict the maximum number of knots to $\sqrt{n}/(2^{d-1})$ for the $d^{th}$ degree basis functions. These hyperparameters can be set through the \texttt{hal9001} package~\cite{coyle2021hal9001-rpkg, HejaziHSHAS2020}. We make the decisions on hyperparameters based on two factors: they can help form a rich model with complex interaction terms and the computing time is acceptable. To make it more rigorous, a cross-validation-based tuning procedure can be considered in future practice.

In the Appendix, we provide a list showing the variables included in the $Q$ models after undersmoothing (Table \ref{tab:variable_list}).

\subsubsection{Data Generating Process}

The DGD for each study was based upon the following structural causal model (SCM):
\begin{eqnarray*}
W &=& f_W(U_W) \\
A &=& f_A(W,U_A) \\
Y &=& f_Y(W,A,U_Y),
\end{eqnarray*}
where $W$, $A$, and $Y$ are, in time ordering, the confounders, the binary intervention of interest and the outcome, respectively, with the $U$ exogenous independent errors and deterministic functions, $f_\cdot$.  
Specifically, the following steps were taken:
\begin{enumerate}

\item  Covariates $W$ were sampled with replacement from the study data sets with sample size $n$, where $n$ is the size of the original data set.

\item The undersmoothed HAL fit was then used to predict $\mathbb{P}(A = 1 | W)$. The intervention $A$ was then sampled using a binomial distribution with the predicted $\mathbb{P}(A = 1 | W)$. 

\item The outcome $Y$ was then simulated with the undersmoothed HAL fit, using the sampled $W$ and simulated $A$ as input. A mean zero, normal random error was added to the simulated $Y$, using a variance based upon the residual variance of the predicted $Y$ (Namely, $\frac{1}{n}\sum_{k=1}^n(\hat{y}_{\lambda} - y_{k})^2$). 

\end{enumerate}
Note, we could have used other ways of estimating the error distribution in step 4, including density estimation using HAL, but we left this for future studies.

Steps 1 through 3 were repeated 500 times to generate the data sets for each simulation. For each of the study data (Table \ref{tab_study}), we repeated these steps and analyzed the performance of the competing estimators separately by study.

\subsubsection{Target parameter}
Our treatment variable $A$ is binary, and our outcome $Y$ is continuous, indicating a height-to-age Z-score. $W$ represents the measured covariates in each study. The data structure is defined as: $O = (W, A, Y) \sim{\mathbb{P}_{0}} \in \mathcal{M}$ with $n$ independent and identically distributed (i.i.d.) observations $O_1,...,O_n$, where $\mathcal{M}$ denotes the set of possible probability distributions of $\mathbb{P}_{0}$. The target parameter is a feature of $\mathbb{P}_{0}$ that is our quantity of interest \cite{SchulerTMLEJ2017}. We selected as our target parameter the average treatment effect (ATE), or $\Psi^F(\mathbb{P}_{U,X}) = \mathbb{E}_{{U,X}}(Y(1)-Y(0))$, $\mathbb{P}_{U,X} \in \mathcal{M}^F$; where $\mathcal{M}^F$ denotes the collection of possible distributions of $(U,X)$ as described by the SCM, and $Y(a)$ is the outcome for a subject if, possibly contrary to fact, they received nutrition intervention $A=a$. Given we simulated the data based upon on our causal model, under randomization assumption and positivity assumption we can show that this causal parameter is identified by the following statistical estimand \cite{PearlC2009}:
$$\Psi(\mathbb{P}_{0}) = \mathbb{E}_{W,0}[\mathbb{E}_{0}(Y|A = 1, W) - \mathbb{E}_{0}(Y|A = 0, W)]$$
We calculate the true ATE value for each study by first randomly drawing a large number of observations ($N=50000$) from the empirical of $W$ and using:
$$
\psi_{0} = 
\frac{1}{N}\sum_{i=1}^{N}[\mathbb{E}_{0}(Y|A = 1, W) - \mathbb{E}_{0}(Y|A = 0, W)]
$$
where we define the $\mathbb{E}_{0}(Y|A = 1, W)$ and $\mathbb{E}_{0}(Y|A = 0, W)$ term using the fitted undersmooth HAL model.  Note that our simulation process insures the randomization assumption is true and there is no asymptotic violation of the positivity assumption.  However, there can be practical violations of positivity (close to 0 or 1 estimated probabilities of getting treatment for some observations given the $W$) which can deferentially impact estimator performance.

\subsection{The Estimation Problem}

The target parameter depends on the true DGD, $\mathbb{P}_{0}$, through the conditional mean $\bar{Q}_{0}(A,W) = \mathbb{E}_{0}(Y|A,W)$ and the marginal distribution $Q_{W,0} = \mathbb{P}_0(W)$ of $W$, so we can write $\Psi(Q_{0})$, where $Q_{0}=(\bar{Q}_{0},Q_{W,0})$.  Our targeted learning estimation procedure begins with estimating the relevant part $Q_{0}$ of the data-generating distribution $\mathbb{P}_{0}$ needed for evaluating the target parameter \cite{LaanTMLLD2006}.

The two general methods we compare are substitution and estimating equation estimators. Depending on the specific estimator, they can depend on estimators of of the propensity score, $g_0(W) = \mathbb{P}(A=1,W)$, the outcome model, $Q_0(A,W)$, and sometimes both.  We use consistent settings when modeling the outcome and the propensity score via Super Learner (see section \ref{computesection} below for details).  

The estimators we compare are not exhaustive and new methods will be developed, so such studies will continue to be important sources of information for deciding what to do in practice.  We quickly describe the particular estimators compared in our study below.

\subsection{Inverse Probability of Treatment Weighting Estimator}

The inverse probability of treatment weighting (IPTW) is a method that relies on estimates of the conditional probability of treatment given covariates $g(W) = \mathbb{P}(A=1|W)$, referred to as the propensity score \cite{ROSENBAUMCRPSA1983}. 
After it is estimated, the propensity score is used to weight observations such that a simple weighted average is a consistent estimate of the particular causal parameter if the propensity score model is consistent \cite{SchulerTMLEJ2017}. For the ATE (if $g$ were known) the weight is $\frac{A}{g(W)} + \frac{1-A}{1-g(W)}$.

The average treatment effect is then estimated by \cite{AustinMBPWD2015}:
$$\psi_{IPTW,n}= \frac{1}{n}\sum_{i=1}^{n}\Big( \frac{A_{i}}{g_n(W_i)}* Y_{i}\Big) - \frac{1}{n}\sum_{i=1}^{n}\Big( \frac{1-A_{i}}{(1-
g_n(W_i))}*Y_{i}\Big)$$
where $g_n(W)$ is the estimate of the true propensity score ($g_0(W)$).  IPTW is not a double robust estimator, in that its consistency depends on consistent estimation of the propensity score \cite{LaanUMCL2003}. As it is not a substitution estimator, it is not as robust to sparsity \cite{SchulerTMLEJ2017}. However, it is a commonly used estimator of the ATE, and its form and relationship to well-known inverse probability methods in the analysis of survey data make it relatively popular.

We derived statistical inference using a conservative standard error which assumes that $g$ is known (there is an extensive literature on IPTW estimators, but \cite{LaanUMCL2003} is a good reference for technical details).  Specifically, the standard error for this estimator was constructed by multiplying $1/\sqrt{n}$ by the standard deviation of the plug-in resulting influence curve:
$$
Y\Big[\frac{A}{g_n(W)} - \frac{1-A}{1-g_n(W)}\Big] - {\psi}_{IPTW,n}
$$

Since IPTW estimator has many problems such as not
invariant to location transformation of the outcome and suffering from the extreme predictions of $g(W)$ (close to 0 or 1), we use the Hajek/stablized IPTW\cite{ipw_hajek_1971} by normalizing the weights of $Y$ as follows:

$$\psi_{IPTW-Hajek,n}  =  \frac{\sum_{i=1}^{n}\Big(\frac{A_{i}}{g_n(W_i)}* Y_{i}\Big)}{\sum_{i=1}^{n}\Big(\frac{A_{i}}{g_n(W_i)}\Big)} - \frac{\sum_{i=1}^{n}\Big(\frac{1-A_{i}}{1-g_n(W_i)}*Y_{i}\Big)}{\sum_{i=1}^{n}\Big( \frac{1-A_{i}}{1-g_n(W_i)}\Big)} $$

\subsubsection{Cross-Validated Inverse Probability of Treatment Weighting (CV-IPTW) Estimator}

To avoid problems that arise when $g(W)$ is overfit, we also implemented the CV-IPTW estimator by adding another layer of cross-validation when estimating the propensity score \cite{ChernozhukovDDMLF2018}. Specifically, the same SL fitting procedure was implemented on training sets. Then, we use this estimate of $g$ on the corresponding validation sets; as such, we employ a nested cross-validation. In practice, we used the ``Split Sequential SL'' method, an approximation to the nested cross-validation proposed by Coyle \cite{CoyleCCTL2017}, to speed up the estimation while obtaining similar results to standard nested cross-validation. More details on the implementation can be found in section~\ref{computesection} below.

\subsection{Augmented Inverse Probability of Treatment Weighted (A-IPTW) Estimator}

The other estimating equation method included in our study is an augmented version of the IPTW estimator, aptly named the augmented inverse probability of treatment weighted (A-IPTW) estimator \cite{GlynnIAIP2010}. It is a double robust estimator that is consistent for the ATE as long as either the propensity score model ($g_0(W)$) or the outcome regression ($Q_0(A,W)$) is correctly specified. When compared with the IPTW estimator in a Monte Carlo simulation, A-IPTW typically outperformed IPTW with a lower mean squared error when either the propensity score or outcome model was misspecified \cite{GlynnIAIP2010}. 

Intuitively, the A-IPTW improves upon IPTW by fully utilizing the information in the conditioning set of covariates $W$, which contains both information about the probability of treatment and information about the outcome variable \cite{GlynnIAIP2010}. More formal justification comes from the fact that the A-IPTW estimator arises as the solution to the efficient influence curve (a key quantity in semiparametric theory), and thus is locally efficient if both $Q$ and $g$ are correctly specified. 

For the ATE,  A-IPTW estimator solves the mean of the empirical efficient influence curve and can be expressed explicitly for the average treatment effect as follows:
$$\psi_{A-IPTW}= \frac{1}{n}\sum_{i=1}^{n}\Big(\Big[\frac{A_iY_i}{g(W_i)}-\frac{(1-A_i)Y_i}{1-g(W_i)}\Big]-\frac{(A_i-g(W_i))}{g(W_i)(1-g(W_i))}\Big[(1-g(W_i)){\mathbb{E}}(Y_i|A_i=1,W_i)+g(W_i){\mathbb{E}}(Y_i|A_i=0,W_i)\Big]\Big).$$

The standard error for this estimator was constructed by multiplying $1/\sqrt{n}$ by the standard deviation of the plug-in efficient influence curve:
$$
(Y-\bar{Q}_n(A,W))\Big[\frac{A}{g_n(W)} - \frac{1-A}{1-g_n(W)}\Big] + (\bar{Q}_n(1,W) - \bar{Q}_n(0,W)) - {\psi}_{A-IPTW,n}.
$$

\subsubsection{Cross-Validated Augmented Inverse Probability of Treatment Weighted (CV-A-IPTW) Estimator}

Similar to CV-IPTW, to avoid overfitting of the outcome model ($Q$) or propensity score model ($g$), we implemented the CV-A-IPTW estimator by adding another layer of cross-validation when estimating the $Q$ and $g$.  In practice, as discussed above for the IPTW estimator, we used the ``Split Sequential SL'' method proposed by Coyle \cite{CoyleCCTL2017} to speed up the estimation (for more details, see section~\ref{computesection} below).




\subsection{Targeted Maximum Likelihood Estimator (TMLE)}
The targeted maximum likelihood estimator (TMLE) is an augmented substitution estimator that, in context of the ATE, adds a targeting step to the original outcome model fit to optimize the bias-variance trade-off for the parameter of interest \cite{LaanTMLLD2006}. Similar to A-IPTW, TMLE is doubly robust, producing unbiased estimates if either $\bar{Q}_{0}(A,W)$ (i.e. $\mathbb{E}_0(Y|A,W)$) or $g_0(W)$ (i.e. $\mathbb{P}_0(A=1|W)$) is correctly specified. It is asymptotically efficient when both quantities are consistently estimated. As it is a substitution estimator, it is typically more robust to outliers and sparsity than EE estimators \cite{SchulerTMLEJ2017}. A finite sample advantage over estimating equation methods comes from the fact that the estimator respects constraints on the parameter bound, such as ensuring that an estimated probability in the $[0,1]$ range \cite{LaanTMLLD2006}.   

The TMLE, like the A-IPTW estimator, requires preliminary estimates of both $g$ and $Q$. The first step in TMLE is finding an initial estimate of the relevant part $Q_{0}$ of data-generating distribution $\mathbb{P}_{0}$. For all estimators, we use an ensemble machine learning algorithm, the Super Learner (SL) algorithm. This avoids arbitrarily using a single algorithm and ensures that the corresponding fit will be optimal (with respect to the true risk) relative to the candidate algorithms used in the estimation.  Once this initial estimate has been found, TMLE updates the initial fit to make an optimal bias-variance trade-off for the target parameter \cite{LaanTMLLD2006}.

For the ATE, the TMLE first requires $\bar{Q}_{n}(A,W)$, the estimate of the conditional expectation of the outcome given the treatment and covariates $\bar{Q}_{0}(A,W)$ \cite{SchulerTMLEJ2017}.  Next is the targeting step for optimizing the bias-variance trade-off for the parameter of interest. The propensity score ($g_0$) can also be estimated with a flexible algorithm like the Super Learner \cite{vanderLaanSL2007}, and these fits are used to predict the conditional probability of treatment and no treatment for each subject ($g_n(W),1-g_n(W)$). These probabilities are used for updating the initial estimate of the outcome model. This updated estimate is then used to generate potential outcomes for when $A=1$ and $A=0$. Like the G-computation estimator, the TMLE estimate of the ATE is calculated as the mean difference between these pairs \cite{SchulerTMLEJ2017}. 

With the ATE as our target parameter, the Super Learner substitution estimator is \cite{LaanTLCI2011}:
$$\psi_{MLE,n}=\Psi(Q_{n}) = \frac{1}{n}\sum_{i=1}^{n}[\bar{Q}_{n}^{0}(1, W_{i}) - \bar{Q}_{n}^{0}(0, W_{i})]$$
where $Q_n$ is the estimate of $Q_0$ and $\bar{Q}_{n}^{0}(\cdot, W)$ the initial estimate of $\bar{Q}_{0}(\cdot, W)$. 

The next step is to update the estimator above toward the parameter of interest. The targeting process uses $g_{n}$ in a so-called clever covariate to define a one-dimensional model for fluctuating the initial estimator. The clever covariate is defined as:
$$H_{n}^{*}(A,W) = \Big(\frac{I(A=1)}{g_{n}(1|W)}-\frac{I(A=0)}{g_{n}(0|W)}\Big)$$

A simple, one-variable logistic regression is then run for the outcome $Y$ on the clever covariate, using $logit\bar{Q}_{n}^{0}(A,W)$ as the offset to estimate the fluctuation parameter $\epsilon$. This is used for updating the initial estimate $\bar{Q}_{n}^{0}$ into a new estimate $\bar{Q}_{n}^{1}$ as follows:
$$logit\bar{Q}_{n}^{1}(A,W) = logit\bar{Q}_{n}^{0}(A,W) + \epsilon_{n}H_{n}^{*}(A,W)$$
where $\epsilon_{n}$ is the estimate of $\epsilon$.

The updated fit is used to calculate the expected outcome under $A=1$ ($\bar{Q}_{n}^{1}(1,W)$) and $A=0$ ($\bar{Q}_{n}^{1}(0,W)$) for all subjects. These estimates are then plugged into the following equation for the final TMLE estimate of the ATE:

$$\psi_{TMLE,n}=\Psi(Q_{n}^{*}) = \frac{1}{n}\sum_{i=1}^{n}[\bar{Q}_{n}^{1}(1, W_{i}) - \bar{Q}_{n}^{1}(0, W_{i})]$$

 
 The fitting of both the $Q$ and $g$ models to the entire data set for the substitution estimator requires entropy assumptions on the fits and underlying true models.  It is possible to violate this assumption by an overfit of the models of the DGD, and this can occur even when cross-validation is used to choose the resulting fits (though, this helps tremendously).  One can generalize both the estimating equation approach and TMLE to estimators that do not need these entropy assumptions by inclusion of an additional layer of cross-validation. This has also been described as double-machine learning in the context of estimating equations \cite{ChernozhukovDDMLF2018}, though it had previously been proposed as a way of robustifying the TMLE \cite{ZhengATCTN2010,LaanTLCI2011}.

The standard error estimate for TMLE can be constructed by multiplying $1/\sqrt{n}$ by the standard deviation of the plug-in efficient influence curve:
$$
(Y-\bar{Q}_n(A,W))\Big[\frac{A}{g_n(W)} - \frac{1-A}{1-g_n(W)}\Big] + (\bar{Q}_n(1,W) - \bar{Q}_n(0,W)) - {\psi}_{TMLE, n}
$$

\subsubsection{Cross-Validated Targeted Maximum Likelihood Estimation (CV-TMLE)}

Though TMLE is a doubly robust and efficient estimator, its performance suffers when the initial estimator is too adaptive \cite{LaanTLCI2011}. Intuitively, if the initial estimator of $Q$ is overfit,  there is not realistic residual variation left for the targeting step and the update is unable to reduce residual bias.

To address these shortcomings of TMLE, cross-validated targeted maximum likelihood estimation (CV-TMLE) was developed \cite{ZhengATCTN2010}. This modified implementation of TMLE utilizes 10-fold cross-validation for the initial estimator to make TMLE more robust in its bias reduction step. The result is that one has greater leeway to use adaptive methods to estimate components of the DGD while keeping realistic residual variation in the validation sample.

Whereas CV-TMLE can add robustness by making the estimator consistent in a larger statistical model, there is still another way for finite sample performance issues to enter estimation.  Specifically, if the data suffers from a lack of experimentation such that $g_n(W)$ gets too close to 0 or 1, then the estimator can begin to suffer from the unstable inverse weighting in the targeting step, a violation ``positivity''.  There are simple methods to avoid this, by choosing a fixed truncation point, such as truncating the estimate of $g$:  $g^*_n=max(min(1-\delta,g_n),\delta)$, for some small $\delta$ (typical value is $\delta=0.025)$.  However, there exists a more sophisticated method that does a type of model selection in estimating the $g$ model which prevents the update from hurting the fit of the $Q$ model.  This is an area of active research and several collaborative-TMLE (C-TMLE) estimators have been proposed, including adaptive selection of the truncation level $\delta$~\cite{LaanTMLLD2006,JuAPSTJ2019a}.

\subsubsection{Collaborative Targeted Maximum Likelihood Estimation (C-TMLE)}

Collaborative Targeted Maximum Likelihood Estimation (C-TMLE) is an extension of TMLE. In the version used for estimation in this study, it applies variable/model selection for nuisance parameter (e.g. the propensity score) estimation in a ``collaborative'' way, by directly optimizing the empirical metric on the causal estimator~\cite{vanderLaanCDRTM2010}.  In this case, we used the  
original  C-TMLE proposed by van der Laan and Gruber~\cite{vanderLaanCDRTM2010}, which is also called ``the greedy C-TMLE algorithm''. It consists of two major steps: first, a sequence of candidate estimators of the nuisance parameter is constructed from a greedy forward stepwise selection procedure; second, cross-validation is used to select the candidate from this sequence which minimizes a criterion that incorporates a measure of bias and variance with respect to the targeted parameter~\cite{vanderLaanCDRTM2010}. More recent development on C-TMLE includes scalable variable-selection C-TMLE~\cite{JuSCTLF2019} and glmnet-C-TMLE algorithm~\cite{JuCLCPA2019}, which might have improved computational efficiency in high-dimensional setting.

\subsection{Computation}
\label{computesection}
Our simulation study was coded in the statistical programming language R\cite{R_general}. We used \texttt{hal9001}\cite{coyle2021hal9001-rpkg, HejaziHSHAS2020} and \texttt{glmnet}\cite{FriedmanRPGL2010} packages to generate the data via undersmoothed HAL. We used \texttt{sl3}\cite{coyle2021sl3}, \texttt{tmle3}\cite{coyle2021tmle3-rpkg} and \texttt{ctmle}\cite{JuGruber_ctmle} packages to implement each of the estimators described above. To estimate the propensity score and the conditional expectation of the outcome, linear models, mean, \texttt{GAM} (general additive models)~\cite{HastieGAMA1986}, \texttt{ranger} (random forest)\cite{WrightRFIRM2017}, \texttt{glmnet} (lasso), and \texttt{XGBoost}\cite{ChenXSTBA2016} with different tuning parameters were used to form the SL library. For ``Study 9'', we dropped \texttt{GAM} and \texttt{ranger} from the learner library to improve the computational efficiency. Ten-fold cross-validation was chosen by default of \texttt{sl3} package for every SL fit. We used logistic regression meta-learner for propensity scores, and non-negative least squares meta-learner for estimating conditional expectation of the outcome. We truncated the propensity score estimates $g_n(W)$ between $[0.025, 0.975]$ for all estimators.

Theoretically, when constructing CV-TMLE, CV-IPTW and CV-A-IPTW estimators, we need to implement nested SL by adding one more layer of cross-validation. Namely, we first split the data, then fit the SL model (which itself uses a cross-validation) on the training set and make predictions on the validation set. Then we rotate the roles of the validation set and finally obtain a vector of cross-validated predictions of propensity scores and conditional expectations. As discussed above, we used the ``Split Sequential SL'' method proposed by Coyle \cite{CoyleCCTL2017}.

After we estimated the relevant parts of the DGD separately for each of the data study data using undersmoothed HAL, the resulting fits were used to simulate data 500 times for each of the 10 studies. Details of the implementation, including the code, can be found in the GitHub repository:
\url{https://github.com/HaodongL/realistic_simu.git}

\section{Results}\label{results}

\subsection{Undersmoothed HAL Models and The True Average Treatment Effect} 

We implemented undersmoothed HAL on the real data and used the fitted model to generate sample for each simulation. Details of each model and the resulting true ATE values are presented in Table~\ref{tab_hal}. 

\begingroup\fontsize{9}{10}\selectfont
\begin{table}[!ht]

\caption{\label{tab_hal} Statistics of the HAL fits to the individual studies, including the sample size, dimension, and number of basis functions used for the treatment model ($g$) and the corresponding outcome model ($Q$), the corresponding lambda penalty and the resulting $L_1$ norm.}
\centering
\scalebox{0.9}{
\begin{tabular}[t]{ccccccccc}
\toprule
StudyID & n & p & TrueATE & Model & Undersmoothed & Num.coef. & Lambda & $L_1$ norm\\
\midrule
 &  &  &  & Q & T & 167 & 2.4e+01 & 5.6e-03\\

\multirow{-2}{*}{\centering\arraybackslash 1} & \multirow{-2}{*}{\centering\arraybackslash 418} & \multirow{-2}{*}{\centering\arraybackslash 20} & \multirow{-2}{*}{\centering\arraybackslash -0.0109} & g & T & 180 & 7.2e-01 & 6.2e-02\\
\cmidrule{1-9}
 &  &  &  & Q & T & 1747 & 3.1e-02 & 4.5e-01\\

\multirow{-2}{*}{\centering\arraybackslash 2} & \multirow{-2}{*}{\centering\arraybackslash 4863} & \multirow{-2}{*}{\centering\arraybackslash 26} & \multirow{-2}{*}{\centering\arraybackslash 0.0507} & g & T & 124 & 3.9e-01 & 1.4e-02\\
\cmidrule{1-9}
 &  &  &  & Q & T & 1496 & 3.0e-02 & 1.9e-01\\

\multirow{-2}{*}{\centering\arraybackslash 3} & \multirow{-2}{*}{\centering\arraybackslash 7399} & \multirow{-2}{*}{\centering\arraybackslash 22} & \multirow{-2}{*}{\centering\arraybackslash 0.0007} & g & T & 6 & 2.6e+01 & 1.6e-03\\
\cmidrule{1-9}
 &  &  &  & Q & T & 503 & 2.3e+00 & 2.1e-02\\

\multirow{-2}{*}{\centering\arraybackslash 4} & \multirow{-2}{*}{\centering\arraybackslash 1204} & \multirow{-2}{*}{\centering\arraybackslash 36} & \multirow{-2}{*}{\centering\arraybackslash -0.0468} & g & T & 5 & 3.8e+02 & 2.2e-06\\
\cmidrule{1-9}
 &  &  &  & Q & T & 448 & 4.5e+00 & 6.9e-03\\

\multirow{-2}{*}{\centering\arraybackslash 5} & \multirow{-2}{*}{\centering\arraybackslash 2396} & \multirow{-2}{*}{\centering\arraybackslash 42} & \multirow{-2}{*}{\centering\arraybackslash -0.0136} & g & T & 15 & 1.8e+02 & 9.0e-06\\
\cmidrule{1-9}
 &  &  &  & Q & T & 2724 & 3.9e-01 & 7.6e-02\\

\multirow{-2}{*}{\centering\arraybackslash 6} & \multirow{-2}{*}{\centering\arraybackslash 3265} & \multirow{-2}{*}{\centering\arraybackslash 18} & \multirow{-2}{*}{\centering\arraybackslash 0.2523} & g & T & 497 & 1.1e+00 & 2.4e-02\\
\cmidrule{1-9}
 &  &  &  & Q & T & 2274 & 2.3e-02 & 1.7e+00\\

\multirow{-2}{*}{\centering\arraybackslash 7} & \multirow{-2}{*}{\centering\arraybackslash 1931} & \multirow{-2}{*}{\centering\arraybackslash 38} & \multirow{-2}{*}{\centering\arraybackslash -0.0310} & g & F & 0 & 9.7e+01 & 0.0e+00\\
\cmidrule{1-9}
 &  &  &  & Q & T & 138 & 1.4e+00 & 2.1e-02\\

\multirow{-2}{*}{\centering\arraybackslash 8} & \multirow{-2}{*}{\centering\arraybackslash 840} & \multirow{-2}{*}{\centering\arraybackslash 30} & \multirow{-2}{*}{\centering\arraybackslash -0.0442} & g & F & 0 & 1.1e+02 & 0.0e+00\\
\cmidrule{1-9}
 &  &  &  & Q & T & 3700 & 1.8e-01 & 3.1e-02\\

\multirow{-2}{*}{\centering\arraybackslash 9} & \multirow{-2}{*}{\centering\arraybackslash 27275} & \multirow{-2}{*}{\centering\arraybackslash 42} & \multirow{-2}{*}{\centering\arraybackslash 0.0089} & g & T & 102 & 2.7e+01 & 7.9e-06\\
\cmidrule{1-9}
 &  &  &  & Q & T & 503 & 1.2e+00 & 7.3e-03\\

\multirow{-2}{*}{\centering\arraybackslash 10} & \multirow{-2}{*}{\centering\arraybackslash 5443} & \multirow{-2}{*}{\centering\arraybackslash 35} & \multirow{-2}{*}{\centering\arraybackslash 0.0203} & g & F & 0 & 3.5e+03 & 0.0e+00\\
\bottomrule
\end{tabular}}
\end{table}
\endgroup{}

For Study 7, 8 and 10, the initial HAL fits of $g$ models contain no variables, so one $A$ is randomized as in a clinical trial. Thereby, the undersoomthing process for $g$ model was omitted for these three studies, and the initial HAL models were used instead.  This is not surprising since all ten studies were randomized controlled trials (RCT). Grouping categorical intervention variables into binary variables at data cleaning step might preserve or change the randomization. The remainder of the studies included basis functions in $W$ and so are more akin to observational studies. For only these studies, we also compare the performance of the estimators above with the standard difference-in-means estimates, which is also provides consistent estimators for the ATE for these three data-generating distributions. On the other hand, the counts of non-zero coefficients (``Num.coef.'' in Table~\ref{tab_hal}) in the undersmoothed $Q$ models are large for the remaining studies, and so, regardless of the original treatment mechanism that underlied these studies, these ones do not come from a simple treatment randomization model. The details on the variables included after undersmoothing can be found in Table \ref{tab:variable_list} in the Appendix.

\subsection{Estimators' Performance} 

\begin{figure}[!ht]
\centering\includegraphics[
  width=17cm,
  height=14cm,
  keepaspectratio
] {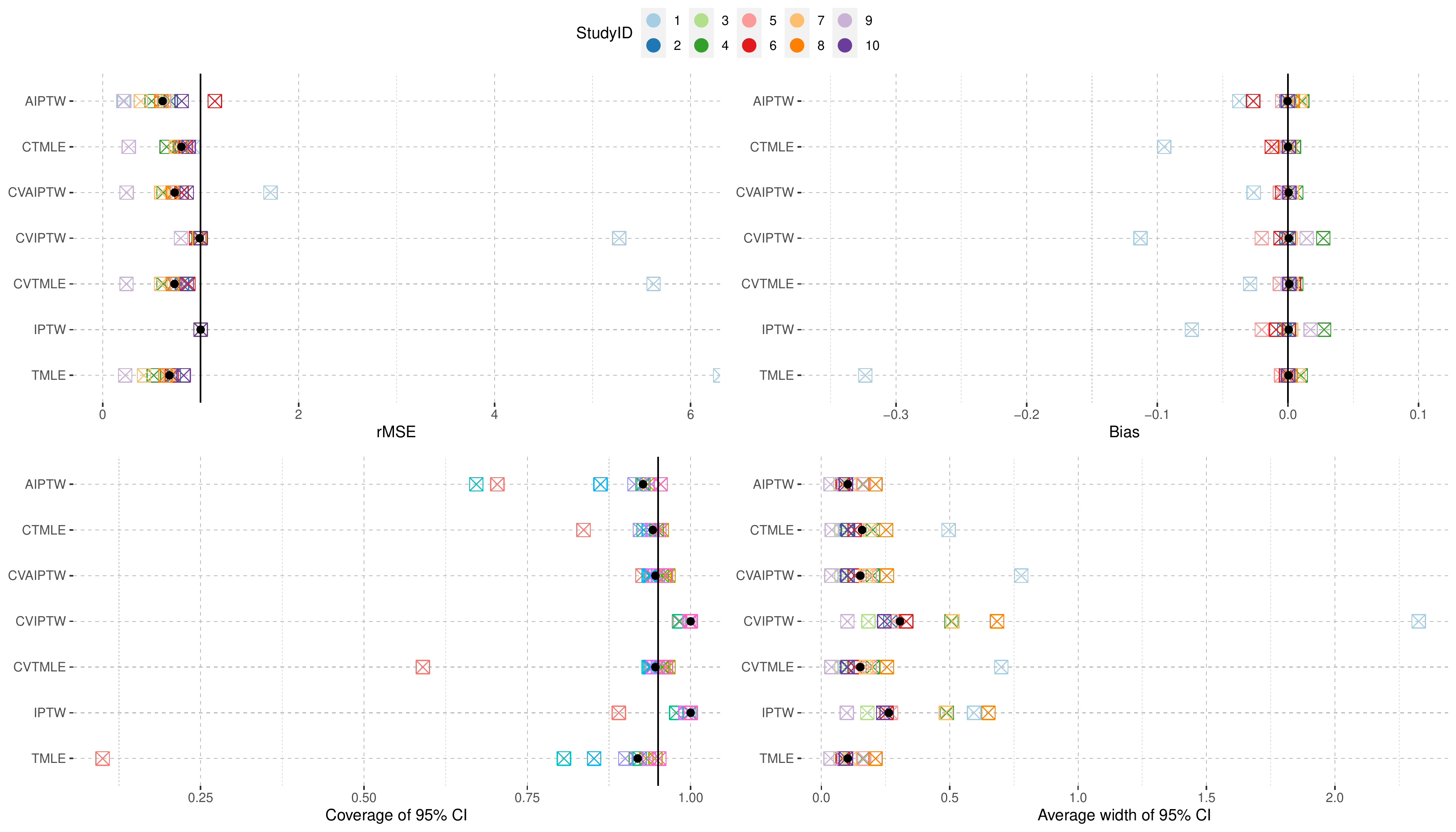}

\raggedright{\scriptsize ~~~~~~~~~~
\textsuperscript{*} The black dots represent the estimator-specific medians across all ten studies.}

{\scriptsize ~~~~~~~~~~
\textsuperscript{*} The reference line is 1 for rMSE, 0 for bias and 0.95 for coverage.}

\raggedright{\scriptsize ~~~~~~~~~~
\textsuperscript{*} The original rMSE value (14.698) of TMLE estimator for Study 1 was truncated at 6.3.}

\caption{\label{fig_perf} \text{Dot plot of the main metrics of performance}}
\end{figure}

The results are shown in Figure \ref{fig_perf} and Table \ref{tab_perf}. Variance dominates bias for all estimators and so contributes overwhelmingly to the mean squared error (MSE) and the relative MSE (rMSE), where rMSE was relative to the IPTW estimator's MSE. Putting aside Study 1 for now, the MSE/rMSE results suggest that the A-IPTW generally is more efficient than the other estimators, the TMLE, CV-TMLE, CV-A-IPTW and C-TMLE with similar MSE to each other, and the IPTW and CV-IPTW having more erratic performance. The bar plots of the main performance metrics in Table  \ref{tab_perf} can be found in the Appendix (see Figure \ref{Figure 3} -  \ref{Figure 10})

The 95\% confidence interval (CI) coverage, however, shows different relative performance (Figure \ref{fig_perf}, Table \ref{tab_perf}). Taking 92.5\% as the lower bound defining consistent coverage, then we can observe that: The CV-A-IPTW consistent coverage for all studies. The CV-TMLE and C-TMLE had consistent coverage for all studies except study 1. The TMLE and A-IPTW had coverage ranging from 90\% to 95\% for most studies. IPTW and CV-IPTW estimates of CI had very conservative coverage (close to 100\%) for most studies.  

To examine more closely issues of CI coverage, we removed the bias introduced by the estimation procedure for the standard error by using the true sample variance of each estimator (i.e. the sample variance of the estimator across 500 simulations) to derive the standard error (``Coverage2'' in Table \ref{tab_perf}). The coverage of this CI is the oracle coverage one would obtain if one is given the true variance. For this measurement, both CV-TMLE and CV-A-IPTW achieved 95\% coverage in all studies, followed by TMLE, C-TMLE, IPTW and CV-IPTW with 95\% coverage for nine studies. A-IPTW had 95\% coverage for eight studies.


\begingroup\fontsize{9}{10}\selectfont

\begin{ThreePartTable}
\begin{TableNotes}
\item[*] ``Variance'' is the true sample variance; ``MSE'' is mean-squared error; ``rMSE'' is the relative (to the IPTW estimator in denominator) mean-squared error; ``Coverage'' is the coverage using 95\% Wald-type confidence intervals (CI) based upon standard error estimates, where ``Coverage2'' uses  the true sample variance; Finally ``CI width'' is the average width of the ``Coverage'' CI's.
\end{TableNotes}
\begin{longtable}[t]{>{\centering\arraybackslash}p{1.1cm}>{\centering\arraybackslash}p{1.1cm}cccc>{\centering\arraybackslash}p{1.7cm}ccc}
\caption{\label{tab_perf}Performance of targeted learning and estimating equation estimators by study within the HAL-based simulations. \textsuperscript{*}}\\
\toprule
Method & StudyID & TrueATE & Variance & Bias & MSE & rMSE & Coverage & Coverage2 & CIwidth\\
\midrule
\endfirsthead
\caption[]{Performance of targeted learning and estimating equation estimators by study within the HAL-based simulations. \textsuperscript{*} \textit{(continued)}}\\
\toprule
Method & StudyID & TrueATE & Variance & Bias & MSE & rMSE & Coverage & Coverage2 & CIwidth\\
\midrule
\endhead

\endfoot
\bottomrule
\insertTableNotes
\endlastfoot
 & 1 & -0.0109 & 0.0056 & -0.0373 & 0.0070 & \includegraphics[width=0.67in, height=0.17in]{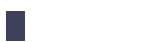} & 0.704 & 0.912 & 0.1658\\
\nopagebreak
 & 2 & 0.0507 & 0.0005 & -0.0012 & 0.0005 & \includegraphics[width=0.67in, height=0.17in]{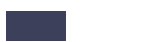} & 0.934 & 0.958 & 0.0849\\
\nopagebreak
 & 3 & 0.0007 & 0.0003 & 0.0007 & 0.0003 & \includegraphics[width=0.67in, height=0.17in]{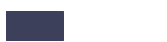} & 0.954 & 0.948 & 0.0737\\
\nopagebreak
 & 4 & -0.0468 & 0.0019 & 0.0109 & 0.0020 & \includegraphics[width=0.67in, height=0.17in]{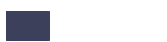} & 0.928 & 0.950 & 0.1612\\
\nopagebreak
 & 5 & -0.0136 & 0.0020 & -0.0046 & 0.0020 & \includegraphics[width=0.67in, height=0.17in]{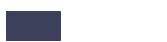} & 0.926 & 0.952 & 0.1640\\
\nopagebreak
 & 6 & 0.2523 & 0.0010 & -0.0266 & 0.0017 & \includegraphics[width=0.67in, height=0.17in]{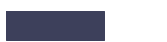} & 0.672 & 0.868 & 0.0829\\
\nopagebreak
 & 7 & -0.0310 & 0.0012 & 0.0093 & 0.0013 & \includegraphics[width=0.67in, height=0.17in]{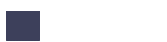} & 0.862 & 0.938 & 0.1098\\
\nopagebreak
 & 8 & -0.0442 & 0.0037 & 0.0037 & 0.0037 & \includegraphics[width=0.67in, height=0.17in]{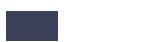} & 0.914 & 0.952 & 0.2112\\
\nopagebreak
 & 9 & 0.0089 & 0.0001 & -0.0005 & 0.0001 & \includegraphics[width=0.67in, height=0.17in]{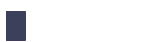} & 0.940 & 0.948 & 0.0362\\
\nopagebreak
\multirow{-10}{1cm}{\centering\arraybackslash A-IPTW} & 10 & 0.0203 & 0.0006 & -0.0001 & 0.0006 & \includegraphics[width=0.67in, height=0.17in]{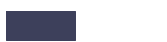} & 0.954 & 0.954 & 0.0961\\
\cmidrule{1-10}\pagebreak[0]
 & 1 & -0.0109 & 0.0219 & -0.0947 & 0.0309 & \includegraphics[width=0.67in, height=0.17in]{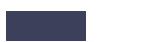} & 0.836 & 0.890 & 0.4956\\
\nopagebreak
 & 2 & 0.0507 & 0.0006 & 0.0016 & 0.0006 & \includegraphics[width=0.67in, height=0.17in]{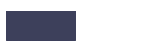} & 0.956 & 0.954 & 0.0993\\
\nopagebreak
 & 3 & 0.0007 & 0.0004 & 0.0018 & 0.0004 & \includegraphics[width=0.67in, height=0.17in]{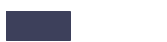} & 0.948 & 0.948 & 0.0782\\
\nopagebreak
 & 4 & -0.0468 & 0.0026 & 0.0046 & 0.0026 & \includegraphics[width=0.67in, height=0.17in]{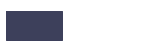} & 0.948 & 0.950 & 0.2005\\
\nopagebreak
 & 5 & -0.0136 & 0.0027 & -0.0087 & 0.0027 & \includegraphics[width=0.67in, height=0.17in]{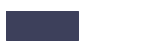} & 0.928 & 0.950 & 0.1882\\
\nopagebreak
 & 6 & 0.2523 & 0.0011 & -0.0124 & 0.0012 & \includegraphics[width=0.67in, height=0.17in]{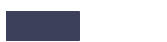} & 0.942 & 0.944 & 0.1295\\
\nopagebreak
 & 7 & -0.0310 & 0.0025 & 0.0012 & 0.0025 & \includegraphics[width=0.67in, height=0.17in]{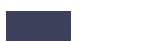} & 0.936 & 0.948 & 0.1875\\
\nopagebreak
 & 8 & -0.0442 & 0.0049 & -0.0014 & 0.0049 & \includegraphics[width=0.67in, height=0.17in]{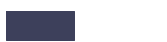} & 0.922 & 0.960 & 0.2524\\
\nopagebreak
 & 9 & 0.0089 & 0.0001 & -0.0008 & 0.0001 & \includegraphics[width=0.67in, height=0.17in]{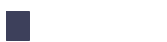} & 0.942 & 0.950 & 0.0409\\
\nopagebreak
\multirow{-10}{1cm}{\centering\arraybackslash C-TMLE} & 10 & 0.0203 & 0.0006 & 0.0007 & 0.0006 & \includegraphics[width=0.67in, height=0.17in]{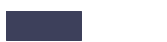} & 0.952 & 0.940 & 0.1037\\
\cmidrule{1-10}\pagebreak[0]
 & 1 & -0.0109 & 0.0565 & -0.0262 & 0.0572 & \includegraphics[width=0.67in, height=0.17in]{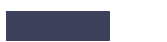} & 0.926 & 0.954 & 0.7789\\
\nopagebreak
 & 2 & 0.0507 & 0.0006 & 0.0031 & 0.0006 & \includegraphics[width=0.67in, height=0.17in]{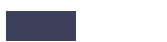} & 0.960 & 0.954 & 0.0985\\
\nopagebreak
 & 3 & 0.0007 & 0.0004 & 0.0009 & 0.0004 & \includegraphics[width=0.67in, height=0.17in]{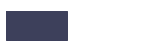} & 0.966 & 0.950 & 0.0793\\
\nopagebreak
 & 4 & -0.0468 & 0.0025 & 0.0063 & 0.0025 & \includegraphics[width=0.67in, height=0.17in]{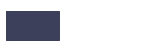} & 0.956 & 0.948 & 0.2008\\
\nopagebreak
 & 5 & -0.0136 & 0.0024 & -0.0062 & 0.0024 & \includegraphics[width=0.67in, height=0.17in]{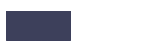} & 0.940 & 0.946 & 0.1881\\
\nopagebreak
 & 6 & 0.2523 & 0.0012 & -0.0045 & 0.0012 & \includegraphics[width=0.67in, height=0.17in]{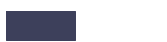} & 0.938 & 0.944 & 0.1301\\
\nopagebreak
 & 7 & -0.0310 & 0.0020 & 0.0030 & 0.0020 & \includegraphics[width=0.67in, height=0.17in]{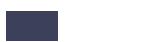} & 0.936 & 0.940 & 0.1737\\
\nopagebreak
 & 8 & -0.0442 & 0.0045 & 0.0000 & 0.0045 & \includegraphics[width=0.67in, height=0.17in]{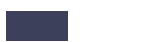} & 0.950 & 0.952 & 0.2553\\
\nopagebreak
 & 9 & 0.0089 & 0.0001 & -0.0002 & 0.0001 & \includegraphics[width=0.67in, height=0.17in]{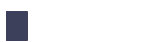} & 0.942 & 0.948 & 0.0394\\
\nopagebreak
\multirow{-10}{1cm}{\centering\arraybackslash CV-A-IPTW} & 10 & 0.0203 & 0.0006 & 0.0011 & 0.0006 & \includegraphics[width=0.67in, height=0.17in]{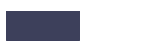} & 0.962 & 0.944 & 0.1026\\
\cmidrule{1-10}\pagebreak[0]
 & 1 & -0.0109 & 0.1632 & -0.1129 & 0.1759 & \includegraphics[width=0.67in, height=0.17in]{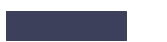} & 0.984 & 0.944 & 2.3263\\
\nopagebreak
 & 2 & 0.0507 & 0.0008 & -0.0012 & 0.0008 & \includegraphics[width=0.67in, height=0.17in]{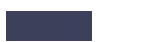} & 1.000 & 0.948 & 0.2686\\
\nopagebreak
 & 3 & 0.0007 & 0.0005 & 0.0020 & 0.0005 & \includegraphics[width=0.67in, height=0.17in]{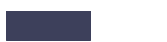} & 1.000 & 0.954 & 0.1831\\
\nopagebreak
 & 4 & -0.0468 & 0.0032 & 0.0270 & 0.0040 & \includegraphics[width=0.67in, height=0.17in]{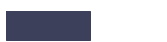} & 1.000 & 0.936 & 0.5065\\
\nopagebreak
 & 5 & -0.0136 & 0.0028 & -0.0202 & 0.0032 & \includegraphics[width=0.67in, height=0.17in]{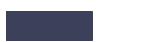} & 0.982 & 0.936 & 0.2817\\
\nopagebreak
 & 6 & 0.2523 & 0.0014 & -0.0057 & 0.0014 & \includegraphics[width=0.67in, height=0.17in]{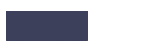} & 1.000 & 0.954 & 0.3305\\
\nopagebreak
 & 7 & -0.0310 & 0.0033 & 0.0017 & 0.0033 & \includegraphics[width=0.67in, height=0.17in]{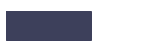} & 1.000 & 0.948 & 0.5111\\
\nopagebreak
 & 8 & -0.0442 & 0.0062 & 0.0006 & 0.0062 & \includegraphics[width=0.67in, height=0.17in]{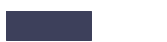} & 1.000 & 0.948 & 0.6842\\
\nopagebreak
 & 9 & 0.0089 & 0.0001 & 0.0143 & 0.0003 & \includegraphics[width=0.67in, height=0.17in]{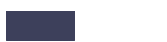} & 0.998 & 0.756 & 0.1016\\
\nopagebreak
\multirow{-10}{1cm}{\centering\arraybackslash CV-IPTW} & 10 & 0.0203 & 0.0007 & 0.0006 & 0.0007 & \includegraphics[width=0.67in, height=0.17in]{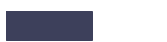} & 1.000 & 0.956 & 0.2451\\
\cmidrule{1-10}\pagebreak[0]
 & 1 & -0.0109 & 0.1868 & -0.0291 & 0.1876 & \includegraphics[width=0.67in, height=0.17in]{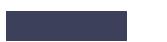} & 0.590 & 0.938 & 0.7006\\
\nopagebreak
 & 2 & 0.0507 & 0.0006 & 0.0031 & 0.0006 & \includegraphics[width=0.67in, height=0.17in]{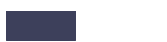} & 0.958 & 0.966 & 0.0985\\
\nopagebreak
 & 3 & 0.0007 & 0.0004 & 0.0009 & 0.0004 & \includegraphics[width=0.67in, height=0.17in]{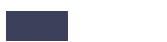} & 0.966 & 0.958 & 0.0793\\
\nopagebreak
 & 4 & -0.0468 & 0.0025 & 0.0064 & 0.0025 & \includegraphics[width=0.67in, height=0.17in]{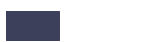} & 0.956 & 0.954 & 0.2008\\
\nopagebreak
 & 5 & -0.0136 & 0.0024 & -0.0063 & 0.0024 & \includegraphics[width=0.67in, height=0.17in]{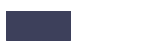} & 0.940 & 0.958 & 0.1881\\
\nopagebreak
 & 6 & 0.2523 & 0.0013 & 0.0046 & 0.0013 & \includegraphics[width=0.67in, height=0.17in]{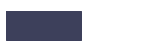} & 0.936 & 0.958 & 0.1301\\
\nopagebreak
 & 7 & -0.0310 & 0.0020 & 0.0030 & 0.0020 & \includegraphics[width=0.67in, height=0.17in]{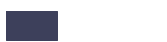} & 0.938 & 0.946 & 0.1737\\
\nopagebreak
 & 8 & -0.0442 & 0.0044 & 0.0000 & 0.0044 & \includegraphics[width=0.67in, height=0.17in]{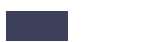} & 0.950 & 0.964 & 0.2551\\
\nopagebreak
 & 9 & 0.0089 & 0.0001 & -0.0002 & 0.0001 & \includegraphics[width=0.67in, height=0.17in]{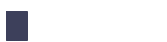} & 0.942 & 0.952 & 0.0394\\
\nopagebreak
\multirow{-10}{1cm}{\centering\arraybackslash CV-TMLE} & 10 & 0.0203 & 0.0006 & 0.0011 & 0.0006 & \includegraphics[width=0.67in, height=0.17in]{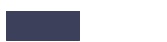} & 0.962 & 0.956 & 0.1026\\
\cmidrule{1-10}\pagebreak[0]
 & 1 & -0.0109 & 0.0280 & -0.0736 & 0.0334 & \includegraphics[width=0.67in, height=0.17in]{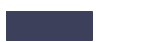} & 0.890 & 0.932 & 0.5945\\
\nopagebreak
 & 2 & 0.0507 & 0.0008 & -0.0028 & 0.0008 & \includegraphics[width=0.67in, height=0.17in]{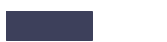} & 1.000 & 0.948 & 0.2544\\
\nopagebreak
 & 3 & 0.0007 & 0.0005 & 0.0022 & 0.0005 & \includegraphics[width=0.67in, height=0.17in]{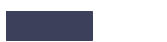} & 1.000 & 0.954 & 0.1789\\
\nopagebreak
 & 4 & -0.0468 & 0.0033 & 0.0276 & 0.0040 & \includegraphics[width=0.67in, height=0.17in]{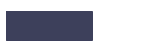} & 1.000 & 0.926 & 0.4889\\
\nopagebreak
 & 5 & -0.0136 & 0.0028 & -0.0201 & 0.0032 & \includegraphics[width=0.67in, height=0.17in]{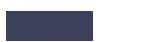} & 0.978 & 0.940 & 0.2712\\
\nopagebreak
 & 6 & 0.2523 & 0.0014 & -0.0091 & 0.0015 & \includegraphics[width=0.67in, height=0.17in]{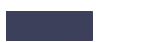} & 0.998 & 0.942 & 0.2536\\
\nopagebreak
 & 7 & -0.0310 & 0.0033 & 0.0023 & 0.0033 & \includegraphics[width=0.67in, height=0.17in]{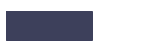} & 1.000 & 0.946 & 0.4838\\
\nopagebreak
 & 8 & -0.0442 & 0.0062 & 0.0003 & 0.0062 & \includegraphics[width=0.67in, height=0.17in]{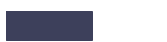} & 1.000 & 0.950 & 0.6504\\
\nopagebreak
 & 9 & 0.0089 & 0.0001 & 0.0172 & 0.0004 & \includegraphics[width=0.67in, height=0.17in]{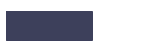} & 0.992 & 0.684 & 0.0990\\
\nopagebreak
\multirow{-10}{1cm}{\centering\arraybackslash IPTW} & 10 & 0.0203 & 0.0007 & 0.0007 & 0.0007 & \includegraphics[width=0.67in, height=0.17in]{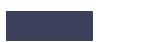} & 1.000 & 0.958 & 0.2410\\
\cmidrule{1-10}\pagebreak[0]
 & 1 & -0.0109 & 0.3860 & -0.3235 & 0.4906 & \includegraphics[width=0.67in, height=0.17in]{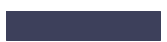} & 0.100 & 0.920 & 0.1681\\
\nopagebreak
 & 2 & 0.0507 & 0.0006 & 0.0005 & 0.0006 & \includegraphics[width=0.67in, height=0.17in]{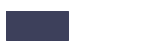} & 0.932 & 0.960 & 0.0849\\
\nopagebreak
 & 3 & 0.0007 & 0.0004 & 0.0007 & 0.0004 & \includegraphics[width=0.67in, height=0.17in]{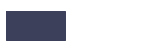} & 0.946 & 0.948 & 0.0737\\
\nopagebreak
 & 4 & -0.0468 & 0.0020 & 0.0099 & 0.0021 & \includegraphics[width=0.67in, height=0.17in]{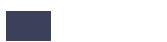} & 0.916 & 0.950 & 0.1611\\
\nopagebreak
 & 5 & -0.0136 & 0.0022 & -0.0052 & 0.0022 & \includegraphics[width=0.67in, height=0.17in]{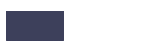} & 0.922 & 0.948 & 0.1640\\
\nopagebreak
 & 6 & 0.2523 & 0.0010 & -0.0015 & 0.0010 & \includegraphics[width=0.67in, height=0.17in]{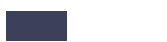} & 0.806 & 0.942 & 0.0828\\
\nopagebreak
 & 7 & -0.0310 & 0.0013 & 0.0079 & 0.0014 & \includegraphics[width=0.67in, height=0.17in]{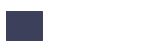} & 0.852 & 0.932 & 0.1098\\
\nopagebreak
 & 8 & -0.0442 & 0.0040 & 0.0020 & 0.0040 & \includegraphics[width=0.67in, height=0.17in]{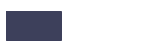} & 0.900 & 0.952 & 0.2111\\
\nopagebreak
 & 9 & 0.0089 & 0.0001 & -0.0003 & 0.0001 & \includegraphics[width=0.67in, height=0.17in]{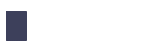} & 0.936 & 0.948 & 0.0362\\
\nopagebreak
\multirow{-10}{1cm}{\centering\arraybackslash TMLE} & 10 & 0.0203 & 0.0006 & 0.0001 & 0.0006 & \includegraphics[width=0.67in, height=0.17in]{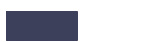} & 0.952 & 0.954 & 0.0961\\*
\end{longtable}
\end{ThreePartTable}
\endgroup{}

The simulations suggest, across 10 realistic data-generating distributions, that CV-A-IPTW, CV-TMLE and C-TMLE has overall relatively good performance in terms of MSE and reliable 95\% coverage.  The A-IPTW estimator had superior MSE-based performance, though the confidence interval coverage was sometimes between 90\% and 95\%. However, plugging in the true standard deviation of the A-IPTW estimator instead of the plug-in influence-curve based one typically used resulted in good coverage. This suggests more robust standard error (SE) estimators could make it a more compelling choice than the empirical performance in these simulations. In addition, CV-A-IPTW can improve the coverage of A-IPTW in most cases, but, due to the estimator being consistent in a bigger model, will have bigger MSE. The results at least show that the CV-TMLE as implemented in the tmle3 package \cite{coyle2021tmle3-rpkg} can provide robust inferences, suggesting using it ``off the shelf'' provides reliable results. In next section, we will discuss situations where even the CV-TMLE under-performed, potentially because of small sample size and related empirical positivity violations \cite{PetersenDRVPF2012}.

\subsection{Exploration on Positivity Violation}
\begin{figure}[!ht]
\centering\includegraphics[
  width=9cm,
  height=4.5cm,
] {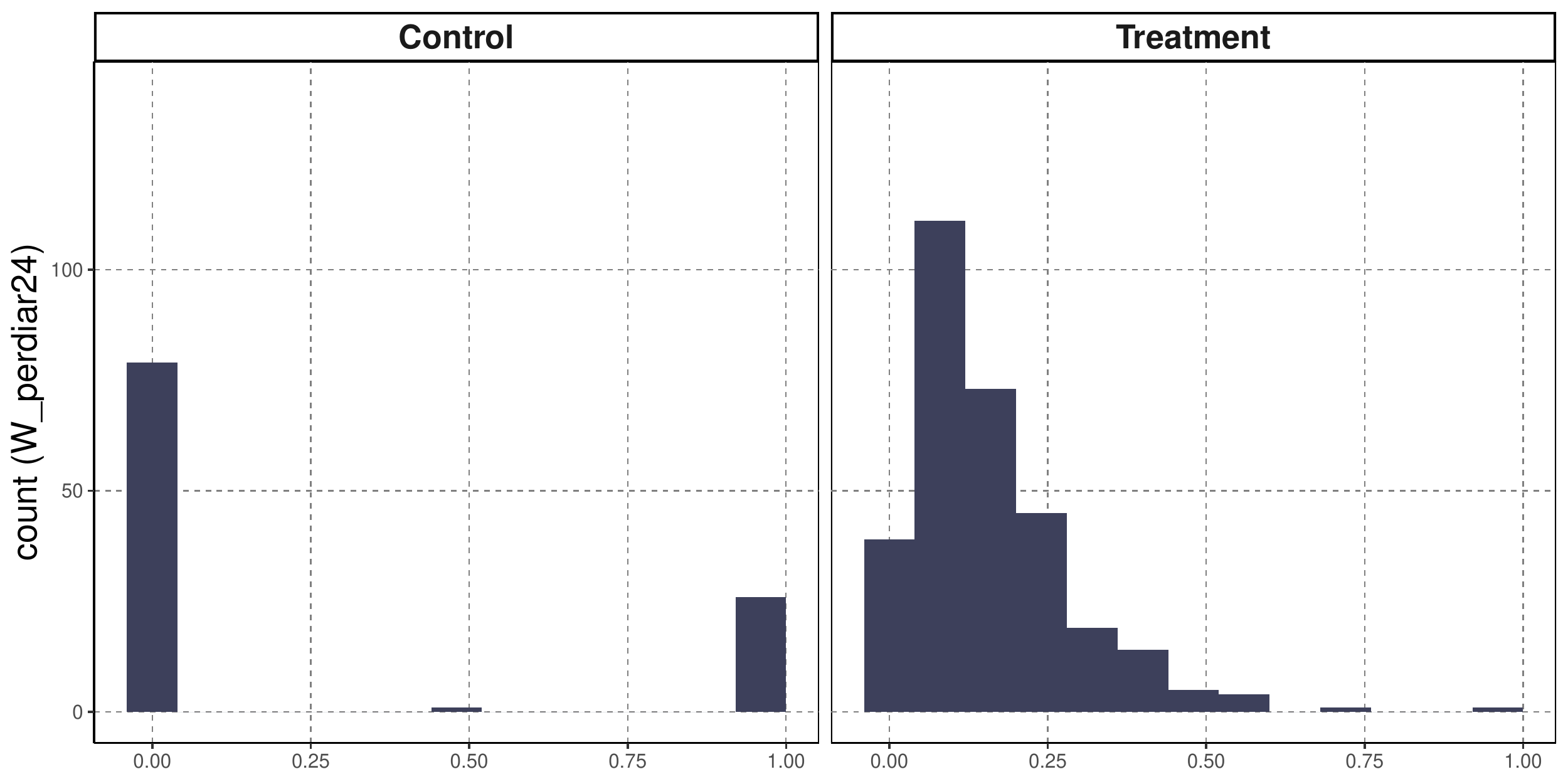}
\caption{\label{fig3} \text{Distributions of \emph{W\_perdiar24} in Study 1 by intervention group}}
\end{figure}

We now consider Study 1, where the TMLE and CV-TMLE had significantly anti-conservative coverage.  In this case, certainly one cause appears to insufficient experimentation of treatment within some covariate groups. 
Specifically, consider Figure \ref{fig3}, which shows the distributions of the adjustment variable, \emph{\text{W\_perdiar24}} in Study 1.  As one can see, there are large differences in the marginal distribution of this covariate; in fact, a fit $g_n$ without smoothing would result in a perfect positivity violation. However, given the variance-bias trade-off resulting in the estimators, it is possible that these empirical violations are smoothed over. A potential consequence of this positivity violation is that the resulting estimator, for the parameter which requires support in the data, will be unstable and biased. 

Table~\ref{tab_dropvar} shows the performance of estimators before and after dropping the variable \emph{\text{W\_perdiar24}} in Study 1. We can observe that all estimators can benefit from removing the problematic variable in terms of higher coverage or lower MSE.

\begingroup\fontsize{9}{10}\selectfont

\begin{table}[!ht]

\caption{\label{tab_dropvar}Estimators' performance with/without \emph{W\_perdiar24} in Study 1 to show the impact of one covariate on performance due to positivity violations. Columns are defined as in table \ref{tab_perf}.} 
\centering
\scalebox{0.9}{
\begin{tabular}[t]{ccccccccc}
\toprule
Method & Dropperdiar & TrueATE & Variance & Bias & MSE & Coverage & Coverage2 & CIwidth\\
\midrule
 & No & -0.0109 & 0.0056 & -0.0373 & 0.0070 & 0.704 & 0.912 & 0.1658\\

\multirow{-2}{*}{\centering\arraybackslash A-IPTW} & Yes & 0.0104 & 0.0084 & -0.0010 & 0.0084 & 0.908 & 0.944 & 0.3117\\
\cmidrule{1-9}
 & No & -0.0109 & 0.0219 & -0.0947 & 0.0309 & 0.836 & 0.890 & 0.4956\\

\multirow{-2}{*}{\centering\arraybackslash C-TMLE} & Yes & 0.0104 & 0.0171 & 0.0020 & 0.0171 & 0.930 & 0.952 & 0.4829\\
\cmidrule{1-9}
 & No & -0.0109 & 0.0565 & -0.0262 & 0.0572 & 0.926 & 0.954 & 0.7789\\

\multirow{-2}{*}{\centering\arraybackslash CV-A-IPTW} & Yes & 0.0104 & 0.0131 & 0.0027 & 0.0131 & 0.950 & 0.940 & 0.4604\\
\cmidrule{1-9}
 & No & -0.0109 & 0.1632 & -0.1129 & 0.1759 & 0.984 & 0.944 & 2.3263\\

\multirow{-2}{*}{\centering\arraybackslash CV-IPTW} & Yes & 0.0104 & 0.0198 & 0.0097 & 0.0199 & 1.000 & 0.948 & 1.3311\\
\cmidrule{1-9}
 & No & -0.0109 & 0.1868 & -0.0291 & 0.1876 & 0.590 & 0.938 & 0.7006\\

\multirow{-2}{*}{\centering\arraybackslash CV-TMLE} & Yes & 0.0104 & 0.0132 & 0.0030 & 0.0132 & 0.950 & 0.946 & 0.4600\\
\cmidrule{1-9}
 & No & -0.0109 & 0.0280 & -0.0736 & 0.0334 & 0.890 & 0.932 & 0.5945\\

\multirow{-2}{*}{\centering\arraybackslash IPTW} & Yes & 0.0104 & 0.0202 & 0.0113 & 0.0203 & 1.000 & 0.948 & 1.2379\\
\cmidrule{1-9}
 & No & -0.0109 & 0.3860 & -0.3235 & 0.4906 & 0.100 & 0.920 & 0.1681\\

\multirow{-2}{*}{\centering\arraybackslash TMLE} & Yes & 0.0104 & 0.0096 & -0.0006 & 0.0096 & 0.878 & 0.942 & 0.3116\\
\bottomrule
\end{tabular}}
\end{table}
\endgroup{}

\subsection{Estimators' Efficiency in Randomized Experiment Setting}

As mentioned in earlier section, the initial HAL models for propensity score include no variables for Study 7, 8 and 10, which leads to randomized experiments in the corresponding simulations. In these cases, we add the ``difference-in-means'' estimator (i.e. $\frac{1}{n_1}\sum_{i=1}^n A_iY_i - \frac{1}{n_0}\sum_{i=1}^n (1-A_i)Y_i$) with its variance estimator proposed by Neyman in 1923~\cite{Splawa-NeymanAPTAN1990}. Table~\ref{table_diffmean} shows that the CV-TMLE and CV-A-IPTW estimators still gain efficiency in the randomized experiments setting.  This is consistent with proposals for using doubly robust estimators of the ATE in randomized trials if there are informative covariates that can increase efficiency over simple, unadjusted estimates \cite{MooreCARTJ2009,tsiatis2008covariate}.

\begingroup\fontsize{9}{10}\selectfont
\begin{table}[!ht]

\caption{\label{table_diffmean}Relative performance of the two CV-estimators with a simple difference in means in the context of the three studies for which treatment was unrelated to covariates (thus equivalent to randomized clinical trial). Columns are defined as in table \ref{tab_perf}.}
\centering
\scalebox{0.9}{
\begin{tabular}[t]{ccccccccc}
\toprule
StudyID & Method & TrueATE & Variance & Bias & MSE & Coverage & Coverage2 & CIwidth\\
\midrule
 & CV-A-IPTW & -0.0310 & 0.0020 & 0.0030 & 0.0020 & 0.936 & 0.940 & 0.1737\\

 & CV-TMLE & -0.0310 & 0.0020 & 0.0030 & 0.0020 & 0.938 & 0.946 & 0.1737\\

\multirow{-3}{*}{\centering\arraybackslash 7} & Diff-in-Mean & -0.0310 & 0.0036 & 0.0021 & 0.0036 & 0.944 & 0.942 & 0.2435\\
\cmidrule{1-9}
 & CV-A-IPTW & -0.0442 & 0.0045 & 0.0000 & 0.0045 & 0.950 & 0.952 & 0.2553\\

 & CV-TMLE & -0.0442 & 0.0044 & 0.0000 & 0.0044 & 0.950 & 0.964 & 0.2551\\

\multirow{-3}{*}{\centering\arraybackslash 8} & Diff-in-Mean & -0.0442 & 0.0068 & -0.0001 & 0.0068 & 0.942 & 0.952 & 0.3115\\
\cmidrule{1-9}
 & CV-A-IPTW & 0.0203 & 0.0006 & 0.0011 & 0.0006 & 0.962 & 0.944 & 0.1026\\

 & CV-TMLE & 0.0203 & 0.0006 & 0.0011 & 0.0006 & 0.962 & 0.956 & 0.1026\\

\multirow{-3}{*}{\centering\arraybackslash 10} & Diff-in-Mean & 0.0203 & 0.0008 & 0.0007 & 0.0008 & 0.964 & 0.948 & 0.1167\\
\bottomrule
\end{tabular}}
\end{table}
\endgroup{}

\section{Conclusion}\label{conclu}

The ultimate goal of studies, such as ours, is to move incrementally towards algorithms that can take information on the design, causal model and known constrains in order to produce a data-adaptively optimized estimator without relying on arbitrary model assumptions. Asymptotic theory can provide guidance on some of the choices, but asymptotic efficiency is not a guarantee for superior performance in finite samples. Thus, simulation studies that are based on realistic DGD's are invaluable for both evaluating estimators and modifying them to increase finite-sample robustness. We provided results supporting the use of a strategically undersmoothed HAL for estimating the relevant components of the DGD in data-driven simulations. Though much remains unresolved, such an approach could be an approach for generating synthetic data \cite{mannino2019real}.

Our results suggest that if accurate inferences are the highest priority, then the CV-A-IPTW, CV-TMLE, and C-TMLE are good choices for providing robust inferences.  Specifically, the results suggest that CV-TMLE might serve as an ``off the shelf'' algorithm given that 1) it is an asymptotically linear estimator; 2) it is consistent in a large class of statistical models; 3) it allows for the use of aggressive ensemble learning, while protecting the final performance of the estimator with an outer layer of cross-validation; 4) its influence-curve-based standard error combined with the well-behaved (normal) distribution of the estimator results in near perfect coverage for all but one of the studies used. Our results also suggest that modifications to the algorithms for other estimators (such as improving the SE estimator for the A-IPTW) would result in an estimator with acceptable CI coverage and relatively low MSE. We also suggest one basis for deciding which estimator to use for particular data is to perform a similar simulation study for the data based upon fitting the undersmoothed HAL to derive the DGD. Then, one could choose to report the results from the estimator that provided the most reliable performance in such a simulation study. Of course, this is itself a form of over-fitting, since it uses the data both for estimator selection and for reporting the results of that estimator applied to the original data. However, it seems better than applying an arbitrary estimator and hoping that the advertised asymptotic performance matches the performance on the data of interest. Finally, our results support the observations that careful use of covariate information can be used to gain efficiency in the randomized experiment setting.

\section*{Acknowledgments}
This research was financially supported by a global development grant (OPP1165144) from the Bill \& Melinda Gates Foundation to the University of California, Berkeley, CA, USA. We would also like to thank the following collaborators on the included cohorts and trials for their contributions to study planning, data collection, and analysis: Muhammad Sharif, Sajjad Kerio, Ms. Urosa, Ms. Alveen, Shahneel Hussain,
Vikas Paudel (Mother and Infant Research Activities), Anthony Costello (University College London), Noel Rouamba, Jean-Bosco Ouédraogo, Leah Prince, Stephen A Vosti, Benjamin Torun, Lindsey M Locks, Christine M McDonald, Roland Kupka, Ronald J Bosch, Rodrick Kisenge, Said Aboud, Molin Wang, Azaduzzaman, Abu Ahmed Shamim, Rezaul Haque, Rolf Klemm, Sucheta Mehra, Maithilee Mitra, Kerry Schulze, Sunita Taneja, Brinda Nayyar, Vandana Suri, Poonam Khokhar, Brinda Nayyar, Poonam Khokhar, Jon E Rohde, Tivendra Kumar, Jose Martines, Maharaj K Bhan, and all other members of the study staffs and field teams. We would also like to thank all study participants and their families for their important contributions. We are grateful to the LCNI5 and iLiNS research teams, participants and people of Lungwena, Namwera, Mangochi and Malindi, our research assistants for their positive attitude, support, and help in all stages of the studies.

In addition, this research used the Savio computational cluster resource provided by the Berkeley Research Computing program at the University of California, Berkeley (supported by the UC Berkeley Chancellor, Vice Chancellor for Research, and Chief Information Officer). The authors would like to further thank the university and the Savio group for providing computational resources.

\subsection*{Author contributions}

Conceptualization: A.E.H., H.L., S.R.  \\
Funding Acquisition: J.M.C., A.E.H., M.J.V., B.F.A. \\
Data curation: A.M., J.B., J.C. \\
Formal analyses: H.L., S.R.\\
Methodology: H.L., S.R., A.E.H., M.J.V. \\
Visualization: H.L., S.R., J.C. \\
Writing – Original Draft Preparation: H.L., S.R. \\
Writing – Review \& Editing: A.E.H., H.L., R.V.P., N.H., I.M., B.F.A., J.B.

\subsection*{Financial disclosure}

None reported. 

\subsection*{Conflict of interest}

The authors declare no potential conflict of interests.

\subsection*{Data availability}

The data used in this analysis was held by Bill \& Melinda Gates Foundation in a repository. The sensitive information contained in the data was still considered theoretically identifiable and can not be released to the public at this point, with the exception of the WASH Benefits trials. We provide the data from "WASH Benefits Bangladesh"\cite{LubyEWQSM2018} (Study 2) and "WASH Benefits Kenya"\cite{StewartEWQSA2018} (Study 3) as example data sets in the GitHub repository:
\url{https://github.com/HaodongL/realistic_simu.git} 

\newpage
\appendix

\section{.  Supplemental figures and tables\label{app1}}

\setcounter{figure}{1} 

\begingroup\fontsize{7}{9}\selectfont

\begin{longtable}[H]{>{\centering\arraybackslash}p{4em}>{\raggedright\arraybackslash}p{12em}>{\raggedright\arraybackslash}p{35em}}
\caption{\label{tab:variable_list}Variables included in the undersmoothed HAL models for the outcome}\\
\toprule
StudyID & AllCovariates & IncludedCovariates\\
\midrule
1 & agedays, sex, month, brthmon, hdlvry, enstunt, enwast, W\_mage, W\_fage, W\_meducyrs, W\_feducyrs, W\_nhh, W\_parity, W\_perdiar24, W\_mhtcm, W\_fhtcm, delta\_W\_mhtcm, delta\_W\_fhtcm, a & agedays*sex*W\_fhtcm, agedays*month*brthmon, agedays*month*enstunt, agedays*month*W\_mage, agedays*month*W\_meducyrs, agedays*month*W\_feducyrs, agedays*month*W\_mhtcm, agedays*brthmon*enstunt, agedays*brthmon*W\_mage, agedays*brthmon*W\_fage, agedays*brthmon*W\_meducyrs, agedays*brthmon*W\_feducyrs, agedays*brthmon*W\_nhh, agedays*brthmon*W\_fhtcm, agedays*hdlvry*W\_mhtcm, agedays*hdlvry*W\_fhtcm, agedays*enstunt*W\_fage, agedays*enstunt*W\_feducyrs, agedays*enstunt*W\_mhtcm, agedays*enstunt*W\_fhtcm, agedays*enwast*W\_feducyrs, agedays*enwast*W\_fhtcm, agedays*W\_mage*W\_fage, agedays*W\_mage*W\_meducyrs, agedays*W\_mage*W\_feducyrs, agedays*W\_mage*W\_nhh, agedays*W\_mage*W\_mhtcm, agedays*W\_mage*W\_fhtcm, agedays*W\_fage*W\_meducyrs, agedays*W\_fage*W\_feducyrs, agedays*W\_fage*W\_parity, agedays*W\_fage*W\_fhtcm, agedays*W\_meducyrs*W\_feducyrs, agedays*W\_meducyrs*W\_fhtcm, agedays*W\_feducyrs*W\_parity, agedays*W\_feducyrs*W\_perdiar24, agedays*W\_feducyrs*W\_mhtcm, agedays*W\_feducyrs*W\_fhtcm, agedays*W\_feducyrs*a, agedays*W\_parity*W\_mhtcm, agedays*W\_parity*W\_fhtcm, agedays*W\_perdiar24*W\_fhtcm, W\_mage*W\_fage*W\_mhtcm, W\_fage*W\_mhtcm*W\_fhtcm, agedays*month*hdlvry, agedays*month*W\_fage, agedays*month*W\_nhh, agedays*month*W\_parity, agedays*month*W\_fhtcm, agedays*brthmon*W\_parity, agedays*brthmon*W\_mhtcm, agedays*W\_mage*W\_parity, agedays*W\_fage*W\_nhh, agedays*W\_fage*W\_mhtcm, agedays*W\_fage*a, agedays*W\_meducyrs*W\_nhh, agedays*W\_nhh*W\_mhtcm, agedays*W\_nhh*W\_fhtcm, agedays*W\_mhtcm*W\_fhtcm, agedays*W\_mhtcm*a\\
\cmidrule{1-3}\pagebreak[0]
2 & sex, month, brthmon, hfoodsec, enstunt, agedays, W\_meducyrs, W\_nhh, W\_mage, W\_mhtcm, W\_mwtkg, W\_mbmi, enwast, impsan, W\_feducyrs, W\_parity, delta\_W\_mage, delta\_W\_mhtcm, delta\_W\_mwtkg, delta\_W\_mbmi, delta\_enwast, delta\_impsan, delta\_W\_feducyrs, delta\_W\_parity, a & sex*agedays, month*agedays, month*W\_mhtcm, brthmon*agedays, brthmon*W\_meducyrs, hfoodsec*agedays, hfoodsec*W\_mage, enstunt*agedays, enstunt*W\_mhtcm, enstunt*W\_mwtkg, enstunt*W\_feducyrs, agedays*W\_meducyrs, agedays*W\_nhh, agedays*W\_mage, agedays*W\_mhtcm, agedays*W\_mwtkg, agedays*W\_mbmi, agedays*enwast, agedays*impsan, agedays*W\_feducyrs, agedays*W\_parity, agedays*delta\_W\_mage, agedays*delta\_W\_mhtcm, agedays*delta\_W\_mwtkg, agedays*delta\_W\_mbmi, agedays*delta\_enwast, agedays*delta\_impsan, agedays*delta\_W\_feducyrs, agedays*delta\_W\_parity, agedays*a, W\_meducyrs*W\_mhtcm, W\_meducyrs*W\_parity, W\_mage*W\_mwtkg, W\_mage*a, W\_mhtcm*W\_mwtkg, W\_mhtcm*W\_mbmi, W\_mhtcm*enwast, W\_mwtkg*enwast, W\_mwtkg*W\_feducyrs, agedays, sex*W\_mage, sex*W\_mhtcm, month*W\_nhh, month*W\_mage, month*W\_mwtkg, month*W\_feducyrs, brthmon*W\_nhh, brthmon*W\_mage, brthmon*W\_mhtcm, brthmon*W\_mwtkg, brthmon*W\_feducyrs, hfoodsec*W\_meducyrs, hfoodsec*W\_mwtkg, W\_meducyrs*W\_mage, W\_meducyrs*W\_mwtkg, W\_meducyrs*W\_feducyrs, W\_nhh*W\_mage, W\_nhh*W\_mhtcm, W\_nhh*W\_mwtkg, W\_nhh*W\_feducyrs, W\_mage*W\_mhtcm, W\_mhtcm*W\_feducyrs, W\_mhtcm*W\_parity, W\_mwtkg*W\_mbmi, W\_mwtkg*W\_parity, W\_mbmi*W\_feducyrs\\
\cmidrule{1-3}\pagebreak[0]
3 & agedays, sex, month, brthmon, enstunt, enwast, cleanck, impfloor, W\_mage, W\_mhtcm, W\_meducyrs, W\_nhh, impsan, delta\_cleanck, delta\_impfloor, delta\_W\_mage, delta\_W\_mhtcm, delta\_W\_meducyrs, delta\_W\_nhh, delta\_impsan, a & agedays*sex, agedays*month, agedays*brthmon, agedays*enstunt, agedays*enwast, agedays*cleanck, agedays*impfloor, agedays*W\_mage, agedays*W\_mhtcm, agedays*W\_meducyrs, agedays*W\_nhh, agedays*impsan, agedays*delta\_cleanck, agedays*delta\_W\_nhh, agedays*delta\_impsan, agedays*a, month*W\_mhtcm, brthmon*W\_mhtcm, enstunt*W\_mage, enstunt*W\_mhtcm, impfloor*W\_mhtcm, W\_mage*W\_mhtcm, W\_mhtcm*W\_meducyrs, W\_mhtcm*W\_nhh, W\_mhtcm*impsan, W\_mhtcm*delta\_W\_mage, W\_mhtcm*delta\_W\_mhtcm, W\_nhh*a, agedays, agedays*delta\_impfloor, agedays*delta\_W\_mage, agedays*delta\_W\_mhtcm, agedays*delta\_W\_meducyrs, month*W\_mage, month*W\_meducyrs, month*W\_nhh, brthmon*W\_mage, brthmon*W\_nhh, W\_mage*W\_meducyrs, W\_mage*W\_nhh, W\_meducyrs*W\_nhh\\
\cmidrule{1-3}\pagebreak[0]
4 & agedays, sex, month, brthmon, enstunt, single, W\_gagebrth, W\_mage, W\_meducyrs, W\_feducyrs, W\_parity, hhwealth\_quart, enwast, vagbrth, hdlvry, earlybf, hfoodsec, W\_birthwt, W\_mhtcm, W\_mwtkg, W\_mbmi, W\_fage, impsan, delta\_enwast, delta\_vagbrth, delta\_hdlvry, delta\_earlybf, delta\_hfoodsec, delta\_W\_birthwt, delta\_W\_mhtcm, delta\_W\_mwtkg, delta\_W\_mbmi, delta\_W\_fage, delta\_impsan, a & agedays*sex, agedays*brthmon, agedays*enstunt, agedays*W\_gagebrth, agedays*W\_birthwt, agedays*W\_mhtcm, agedays*W\_fage, agedays*delta\_hdlvry, agedays*delta\_hfoodsec, agedays*a, sex*W\_birthwt, brthmon*W\_birthwt, enstunt*W\_birthwt, single*W\_birthwt, W\_gagebrth*W\_birthwt, W\_gagebrth*W\_mwtkg, W\_mage*W\_birthwt, W\_meducyrs*W\_birthwt, W\_feducyrs*W\_birthwt, W\_parity*W\_birthwt, hhwealth\_quart*W\_birthwt, enwast*W\_birthwt, vagbrth*W\_birthwt, hdlvry*W\_birthwt, earlybf*W\_birthwt, hfoodsec*W\_birthwt, W\_birthwt*W\_mhtcm, W\_birthwt*W\_mbmi, W\_birthwt*W\_fage, W\_birthwt*impsan, W\_birthwt*delta\_enwast, W\_birthwt*delta\_hdlvry, W\_birthwt*delta\_earlybf, W\_birthwt*delta\_W\_birthwt, agedays*month, agedays*W\_mage, agedays*W\_meducyrs, agedays*W\_feducyrs, agedays*W\_parity, agedays*hhwealth\_quart, agedays*hfoodsec, agedays*W\_mwtkg, agedays*W\_mbmi, month*W\_birthwt, W\_birthwt*W\_mwtkg, W\_birthwt*delta\_hfoodsec, W\_birthwt*delta\_W\_fage\\
\cmidrule{1-3}\pagebreak[0]
5 & agedays, sex, month, brthmon, enstunt, anywast06, enwast, vagbrth, hdlvry, single, nchldlt5, hhwealth\_quart, pers\_wast, W\_gagebrth, W\_birthwt, W\_mage, W\_mhtcm, W\_mwtkg, W\_mbmi, W\_meducyrs, W\_feducyrs, W\_nchldlt5, W\_parity, delta\_enwast, delta\_vagbrth, delta\_hdlvry, delta\_single, delta\_nchldlt5, delta\_hhwealth\_quart, delta\_pers\_wast, delta\_W\_gagebrth, delta\_W\_birthwt, delta\_W\_mage, delta\_W\_mhtcm, delta\_W\_mwtkg, delta\_W\_mbmi, delta\_W\_meducyrs, delta\_W\_feducyrs, delta\_W\_nchldlt5, delta\_W\_parity, a & agedays*W\_gagebrth, agedays*W\_birthwt, agedays*W\_mhtcm, agedays*W\_meducyrs, sex*W\_birthwt, brthmon*W\_birthwt, enstunt*W\_birthwt, anywast06*W\_birthwt, enwast*W\_birthwt, vagbrth*W\_birthwt, nchldlt5*W\_birthwt, hhwealth\_quart*W\_birthwt, pers\_wast*W\_birthwt, W\_gagebrth*W\_birthwt, W\_birthwt*W\_mage, W\_birthwt*W\_mhtcm, W\_birthwt*W\_mwtkg, W\_birthwt*W\_meducyrs, W\_birthwt*W\_feducyrs, W\_birthwt*W\_nchldlt5, W\_birthwt*W\_parity, W\_birthwt*delta\_hdlvry, W\_birthwt*delta\_single, W\_birthwt*delta\_pers\_wast, W\_birthwt*delta\_W\_birthwt, W\_birthwt*delta\_W\_mage, W\_birthwt*delta\_W\_parity, W\_birthwt*a, agedays*month, agedays*brthmon, agedays*W\_mage, agedays*W\_mwtkg, agedays*W\_mbmi, agedays*W\_feducyrs, month*W\_birthwt, W\_gagebrth*W\_mhtcm, W\_birthwt*W\_mbmi, W\_birthwt*delta\_W\_gagebrth\\
\cmidrule{1-3}\pagebreak[0]
6 & agedays, sex, month, brthmon, enstunt, enwast, W\_mage, W\_mhtcm, W\_mwtkg, W\_mbmi, W\_nchldlt5, delta\_W\_mage, delta\_W\_mhtcm, delta\_W\_mwtkg, delta\_W\_mbmi, delta\_W\_nchldlt5, a & agedays*W\_mage, agedays*W\_mhtcm, agedays*W\_mbmi, W\_mhtcm*W\_mwtkg, agedays*sex*month, agedays*sex*brthmon, agedays*sex*W\_mage, agedays*sex*W\_mhtcm, agedays*month*brthmon, agedays*month*enstunt, agedays*month*W\_mage, agedays*month*W\_mhtcm, agedays*month*W\_mwtkg, agedays*month*W\_mbmi, agedays*month*W\_nchldlt5, agedays*month*delta\_W\_nchldlt5, agedays*brthmon*enstunt, agedays*brthmon*W\_mage, agedays*brthmon*W\_mhtcm, agedays*brthmon*W\_mwtkg, agedays*brthmon*W\_mbmi, agedays*brthmon*W\_nchldlt5, agedays*brthmon*delta\_W\_mhtcm, agedays*brthmon*a, agedays*enstunt*W\_mage, agedays*enstunt*W\_mhtcm, agedays*enstunt*W\_mwtkg, agedays*enstunt*W\_mbmi, agedays*enstunt*W\_nchldlt5, agedays*enstunt*a, agedays*enwast*W\_mhtcm, agedays*enwast*W\_mwtkg, agedays*enwast*a, agedays*W\_mage*W\_mhtcm, agedays*W\_mage*W\_mwtkg, agedays*W\_mage*W\_mbmi, agedays*W\_mage*a, agedays*W\_mhtcm*W\_mwtkg, agedays*W\_mhtcm*W\_mbmi, agedays*W\_mhtcm*W\_nchldlt5, agedays*W\_mhtcm*delta\_W\_mage, agedays*W\_mhtcm*delta\_W\_mhtcm, agedays*W\_mhtcm*delta\_W\_mwtkg, agedays*W\_mhtcm*delta\_W\_mbmi, agedays*W\_mhtcm*delta\_W\_nchldlt5, agedays*W\_mhtcm*a, agedays*W\_mwtkg*W\_mbmi, agedays*W\_mwtkg*W\_nchldlt5, agedays*W\_mwtkg*delta\_W\_mage, agedays*W\_mwtkg*delta\_W\_mhtcm, agedays*W\_mwtkg*delta\_W\_mwtkg, agedays*W\_mwtkg*delta\_W\_mbmi, agedays*W\_mwtkg*delta\_W\_nchldlt5, agedays*W\_mbmi*W\_nchldlt5, agedays*W\_nchldlt5*a, sex*brthmon*W\_mhtcm, sex*W\_mage*W\_mwtkg, month*W\_mage*W\_mhtcm, month*W\_mhtcm*W\_mwtkg, brthmon*enstunt*W\_mhtcm, brthmon*W\_mage*W\_mhtcm, brthmon*W\_mage*W\_mwtkg, brthmon*W\_mhtcm*W\_mwtkg, enstunt*W\_mage*W\_mhtcm, enstunt*W\_mhtcm*W\_mwtkg, enwast*W\_mage*W\_mhtcm, enwast*W\_mhtcm*W\_mwtkg, W\_mage*W\_mhtcm*W\_mwtkg, W\_mage*W\_mhtcm*W\_nchldlt5, W\_mage*W\_mwtkg*W\_mbmi, W\_mhtcm*W\_mwtkg*W\_mbmi, W\_mhtcm*W\_mwtkg*W\_nchldlt5, agedays*W\_mwtkg, agedays*sex*enwast, agedays*sex*W\_mwtkg, agedays*sex*W\_mbmi, agedays*sex*W\_nchldlt5, agedays*month*enwast, agedays*month*delta\_W\_mage, agedays*month*delta\_W\_mhtcm, agedays*month*delta\_W\_mwtkg, agedays*month*delta\_W\_mbmi, agedays*month*a, agedays*brthmon*enwast, agedays*enwast*W\_mage, agedays*enwast*W\_mbmi, agedays*enwast*W\_nchldlt5, agedays*W\_mage*W\_nchldlt5, agedays*W\_mwtkg*a, sex*W\_mhtcm*W\_mwtkg, month*brthmon*W\_mwtkg, month*W\_mage*W\_mwtkg, month*W\_mage*W\_mbmi, month*W\_mwtkg*W\_mbmi, brthmon*W\_mwtkg*W\_mbmi, W\_mage*W\_mhtcm*W\_mbmi, W\_mage*W\_mwtkg*W\_nchldlt5\\
\cmidrule{1-3}\pagebreak[0]
7 & agedays, sex, month, brthmon, enstunt, enwast, single, cleanck, hfoodsec, W\_mage, W\_mhtcm, W\_mwtkg, hhwealth\_quart, safeh20, W\_mbmi, W\_fage, W\_meducyrs, W\_feducyrs, W\_nrooms, W\_nchldlt5, impsan, delta\_single, delta\_cleanck, delta\_hfoodsec, delta\_W\_mage, delta\_W\_mhtcm, delta\_W\_mwtkg, delta\_hhwealth\_quart, delta\_safeh20, delta\_W\_mbmi, delta\_W\_fage, delta\_W\_meducyrs, delta\_W\_feducyrs, delta\_W\_nrooms, delta\_W\_nchldlt5, delta\_impsan, a & agedays*sex, agedays*month, agedays*brthmon, agedays*enstunt, agedays*enwast, agedays*single, agedays*cleanck, agedays*hfoodsec, agedays*W\_mage, agedays*W\_mhtcm, agedays*W\_mwtkg, agedays*hhwealth\_quart, agedays*W\_mbmi, agedays*W\_fage, agedays*W\_meducyrs, agedays*W\_feducyrs, agedays*W\_nrooms, agedays*W\_nchldlt5, agedays*impsan, agedays*delta\_single, agedays*delta\_cleanck, agedays*delta\_hfoodsec, agedays*delta\_W\_mage, agedays*delta\_W\_mhtcm, agedays*delta\_W\_mwtkg, agedays*delta\_hhwealth\_quart, agedays*delta\_safeh20, agedays*delta\_W\_mbmi, agedays*delta\_W\_fage, agedays*delta\_W\_meducyrs, agedays*delta\_W\_feducyrs, agedays*delta\_W\_nrooms, agedays*delta\_W\_nchldlt5, agedays*delta\_impsan, agedays*a, sex*month, month*single, month*W\_mhtcm, month*W\_mwtkg, month*hhwealth\_quart, brthmon*enstunt, brthmon*W\_mwtkg, brthmon*safeh20, brthmon*W\_fage, brthmon*W\_feducyrs, brthmon*W\_nrooms, brthmon*W\_nchldlt5, enstunt*W\_mhtcm, enstunt*W\_mbmi, enstunt*W\_fage, enstunt*W\_feducyrs, single*W\_mhtcm, hfoodsec*W\_mhtcm, hfoodsec*hhwealth\_quart, hfoodsec*W\_meducyrs, W\_mage*W\_mhtcm, W\_mage*W\_mwtkg, W\_mage*hhwealth\_quart, W\_mage*W\_fage, W\_mage*W\_meducyrs, W\_mage*W\_nchldlt5, W\_mage*delta\_W\_nrooms, W\_mhtcm*W\_mwtkg, W\_mhtcm*hhwealth\_quart, W\_mhtcm*W\_mbmi, W\_mhtcm*W\_fage, W\_mhtcm*W\_nrooms, W\_mhtcm*W\_nchldlt5, W\_mhtcm*delta\_impsan, W\_mwtkg*hhwealth\_quart, W\_mwtkg*W\_fage, W\_mwtkg*W\_meducyrs, W\_mwtkg*W\_feducyrs, W\_mwtkg*W\_nchldlt5, W\_mwtkg*impsan, W\_mwtkg*delta\_W\_nrooms, W\_mwtkg*delta\_W\_nchldlt5, W\_mwtkg*a, W\_fage*W\_feducyrs, W\_fage*W\_nrooms, W\_meducyrs*W\_nrooms, W\_meducyrs*W\_nchldlt5, W\_feducyrs*delta\_W\_nrooms, W\_feducyrs*a, agedays, agedays*safeh20, sex*W\_mwtkg, sex*W\_fage, month*brthmon, month*W\_mage, month*W\_mbmi, month*W\_fage, month*W\_meducyrs, month*W\_feducyrs, month*W\_nrooms, brthmon*hfoodsec, brthmon*W\_mage, brthmon*W\_mhtcm, brthmon*W\_mbmi, brthmon*W\_meducyrs, enstunt*W\_mage, enstunt*W\_mwtkg, enwast*W\_mage, single*W\_mage, single*W\_mwtkg, hfoodsec*W\_mage, hfoodsec*W\_mwtkg, hfoodsec*W\_fage, W\_mage*W\_mbmi, W\_mage*W\_feducyrs, W\_mage*W\_nrooms, W\_mhtcm*W\_meducyrs, W\_mhtcm*W\_feducyrs, W\_mwtkg*W\_mbmi, W\_mwtkg*W\_nrooms, W\_mwtkg*delta\_impsan, hhwealth\_quart*W\_fage, hhwealth\_quart*W\_meducyrs, hhwealth\_quart*W\_feducyrs, W\_mbmi*W\_fage, W\_mbmi*W\_meducyrs, W\_fage*W\_meducyrs, W\_fage*W\_nchldlt5, W\_meducyrs*W\_feducyrs, W\_feducyrs*W\_nchldlt5\\
\cmidrule{1-3}\pagebreak[0]
8 & agedays, sex, month, brthmon, enstunt, enwast, W\_mage, W\_mhtcm, W\_mwtkg, W\_mbmi, W\_meducyrs, W\_feducyrs, W\_nchldlt5, W\_nhh, impsan, hhwealth\_quart, safeh20, delta\_W\_mage, delta\_W\_mhtcm, delta\_W\_mwtkg, delta\_W\_mbmi, delta\_W\_meducyrs, delta\_W\_feducyrs, delta\_W\_nchldlt5, delta\_W\_nhh, delta\_impsan, delta\_hhwealth\_quart, delta\_safeh20, a & agedays*sex, agedays*month, agedays*brthmon, agedays*enstunt, agedays*enwast, agedays*W\_mage, agedays*W\_mhtcm, agedays*W\_mbmi, agedays*W\_meducyrs, agedays*W\_feducyrs, agedays*W\_nchldlt5, agedays*W\_nhh, agedays*hhwealth\_quart, agedays*delta\_W\_mage, agedays*delta\_W\_feducyrs, agedays*delta\_W\_nchldlt5, W\_mage*W\_feducyrs, W\_mhtcm*W\_mwtkg, agedays*W\_mwtkg, agedays*safeh20, agedays*a\\
\cmidrule{1-3}\pagebreak[0]
9 & agedays, sex, month, brthmon, enstunt, W\_parity, vagbrth, impfloor, earlybf, hfoodsec, enwast, hhwealth\_quart, safeh20, W\_gagebrth, W\_birthwt, W\_birthlen, W\_mage, W\_mhtcm, W\_nrooms, W\_meducyrs, W\_feducyrs, W\_nchldlt5, impsan, delta\_vagbrth, delta\_impfloor, delta\_earlybf, delta\_hfoodsec, delta\_enwast, delta\_hhwealth\_quart, delta\_safeh20, delta\_W\_gagebrth, delta\_W\_birthwt, delta\_W\_birthlen, delta\_W\_mage, delta\_W\_mhtcm, delta\_W\_nrooms, delta\_W\_meducyrs, delta\_W\_feducyrs, delta\_W\_nchldlt5, delta\_impsan, a & agedays*month, agedays*brthmon, agedays*enstunt, agedays*W\_parity, agedays*earlybf, agedays*enwast, agedays*W\_gagebrth, agedays*W\_birthwt, agedays*W\_birthlen, agedays*W\_mage, agedays*W\_mhtcm, agedays*W\_nrooms, agedays*W\_meducyrs, agedays*W\_feducyrs, agedays*W\_nchldlt5, agedays*impsan, agedays*delta\_enwast, agedays*delta\_W\_gagebrth, agedays*delta\_W\_birthwt, agedays*delta\_W\_meducyrs, sex*W\_birthwt, month*W\_birthwt, brthmon*W\_birthwt, enstunt*W\_birthwt, W\_parity*W\_birthwt, vagbrth*W\_birthwt, impfloor*W\_birthwt, earlybf*W\_birthwt, hfoodsec*W\_birthwt, enwast*W\_birthwt, hhwealth\_quart*W\_birthwt, safeh20*W\_birthwt, W\_gagebrth*W\_birthwt, W\_gagebrth*W\_birthlen, W\_gagebrth*W\_mage, W\_gagebrth*W\_mhtcm, W\_birthwt*W\_birthlen, W\_birthwt*W\_mage, W\_birthwt*W\_mhtcm, W\_birthwt*W\_nrooms, W\_birthwt*W\_meducyrs, W\_birthwt*W\_feducyrs, W\_birthwt*W\_nchldlt5, W\_birthwt*delta\_vagbrth, W\_birthwt*delta\_hfoodsec, W\_birthwt*delta\_enwast, W\_birthwt*delta\_W\_birthlen, W\_birthwt*delta\_W\_feducyrs, W\_birthwt*delta\_impsan, W\_birthwt*a, W\_mhtcm*W\_meducyrs, W\_birthwt, agedays*sex, agedays*impfloor, agedays*hfoodsec, agedays*hhwealth\_quart, agedays*delta\_impfloor, agedays*delta\_earlybf, agedays*delta\_hfoodsec, agedays*delta\_hhwealth\_quart, agedays*delta\_W\_birthlen, agedays*delta\_W\_mhtcm, agedays*delta\_W\_feducyrs, agedays*a, month*W\_gagebrth, brthmon*W\_gagebrth, W\_gagebrth*W\_meducyrs, W\_gagebrth*W\_feducyrs, W\_birthwt*impsan, W\_birthwt*delta\_earlybf, W\_birthwt*delta\_W\_gagebrth, W\_birthwt*delta\_W\_mhtcm, W\_birthwt*delta\_W\_meducyrs, W\_mage*W\_feducyrs, W\_mhtcm*W\_feducyrs\\
\cmidrule{1-3}\pagebreak[0]
10 & agedays, sex, month, brthmon, earlybf, enstunt, W\_perdiar24, vagbrth, hdlvry, impfloor, hfoodsec, enwast, hhwealth\_quart, safeh20, W\_birthwt, W\_birthlen, W\_meducyrs, W\_nrooms, W\_feducyrs, impsan, delta\_vagbrth, delta\_hdlvry, delta\_impfloor, delta\_hfoodsec, delta\_enwast, delta\_hhwealth\_quart, delta\_safeh20, delta\_W\_birthwt, delta\_W\_birthlen, delta\_W\_meducyrs, delta\_W\_nrooms, delta\_W\_feducyrs, delta\_impsan, a & agedays*enstunt, agedays*hfoodsec, agedays*W\_birthwt, agedays*W\_birthlen, agedays*W\_meducyrs, agedays*W\_nrooms, agedays*delta\_enwast, sex*W\_birthwt, month*W\_birthwt, brthmon*W\_birthwt, earlybf*W\_birthwt, enstunt*W\_birthwt, vagbrth*W\_birthwt, hdlvry*W\_birthwt, impfloor*W\_birthwt, hhwealth\_quart*W\_birthwt, W\_birthwt*W\_birthlen, W\_birthwt*W\_meducyrs, W\_birthwt*W\_nrooms, W\_birthwt*W\_feducyrs, W\_birthwt*impsan, W\_birthwt*delta\_vagbrth, W\_birthwt*delta\_hfoodsec, W\_birthwt*delta\_W\_feducyrs, W\_birthwt*a, agedays*month, agedays*brthmon, agedays*hhwealth\_quart, agedays*W\_feducyrs, hfoodsec*W\_birthwt, enwast*W\_birthwt, W\_birthwt*delta\_hdlvry\\
\bottomrule
\end{longtable}
\endgroup{}

\setcounter{figure}{0} 
\begin{figure}[!ht]
\centering\includegraphics[
  width=15cm,
  height=10cm,
  keepaspectratio
] {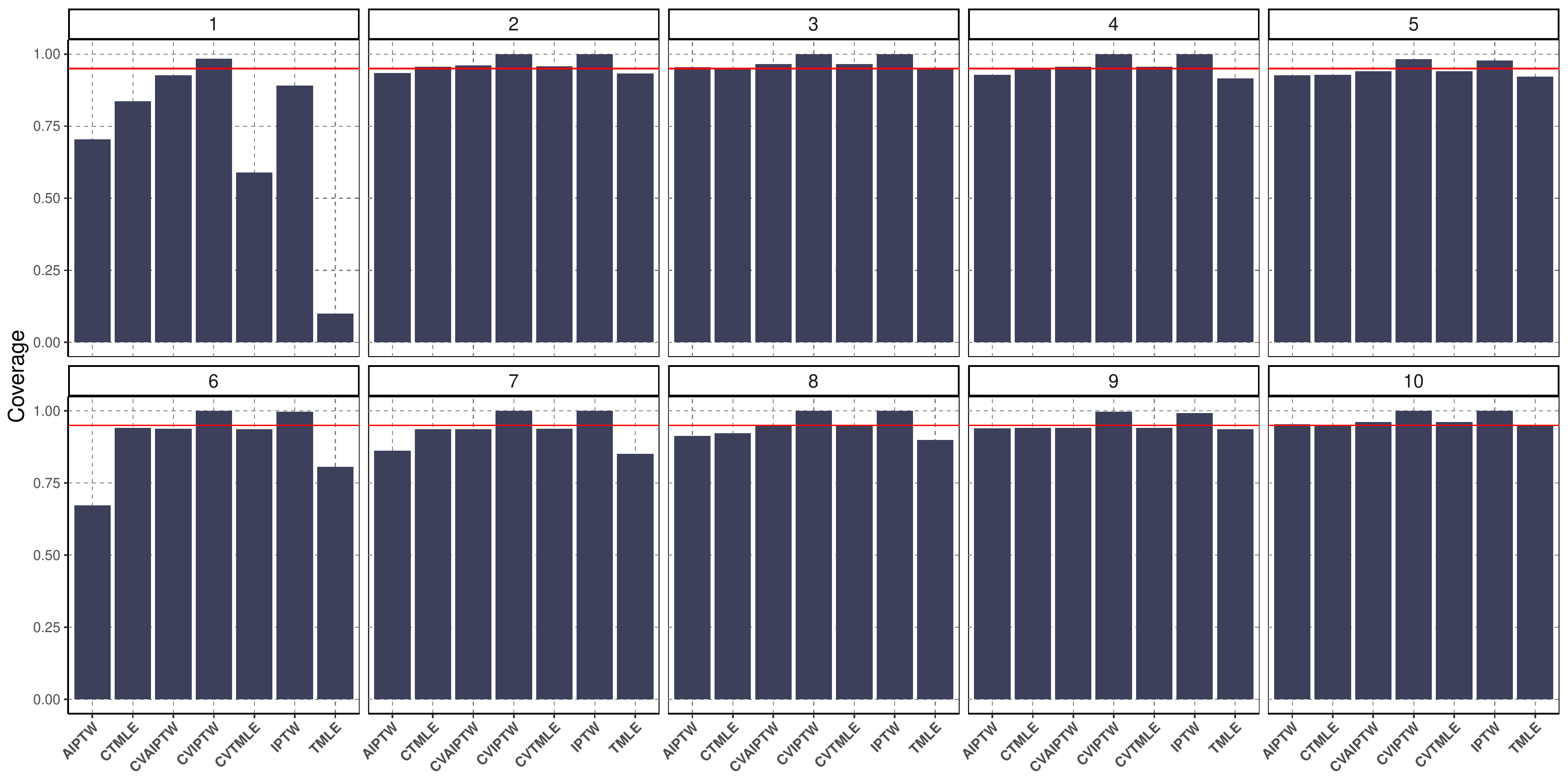}
\caption{\label{Figure 3} \text{Coverage of confidence intervals in ten studies}}
\end{figure}

\begin{figure}[!ht]
\centering\includegraphics[
  width=15cm,
  height=10cm,
  keepaspectratio
] {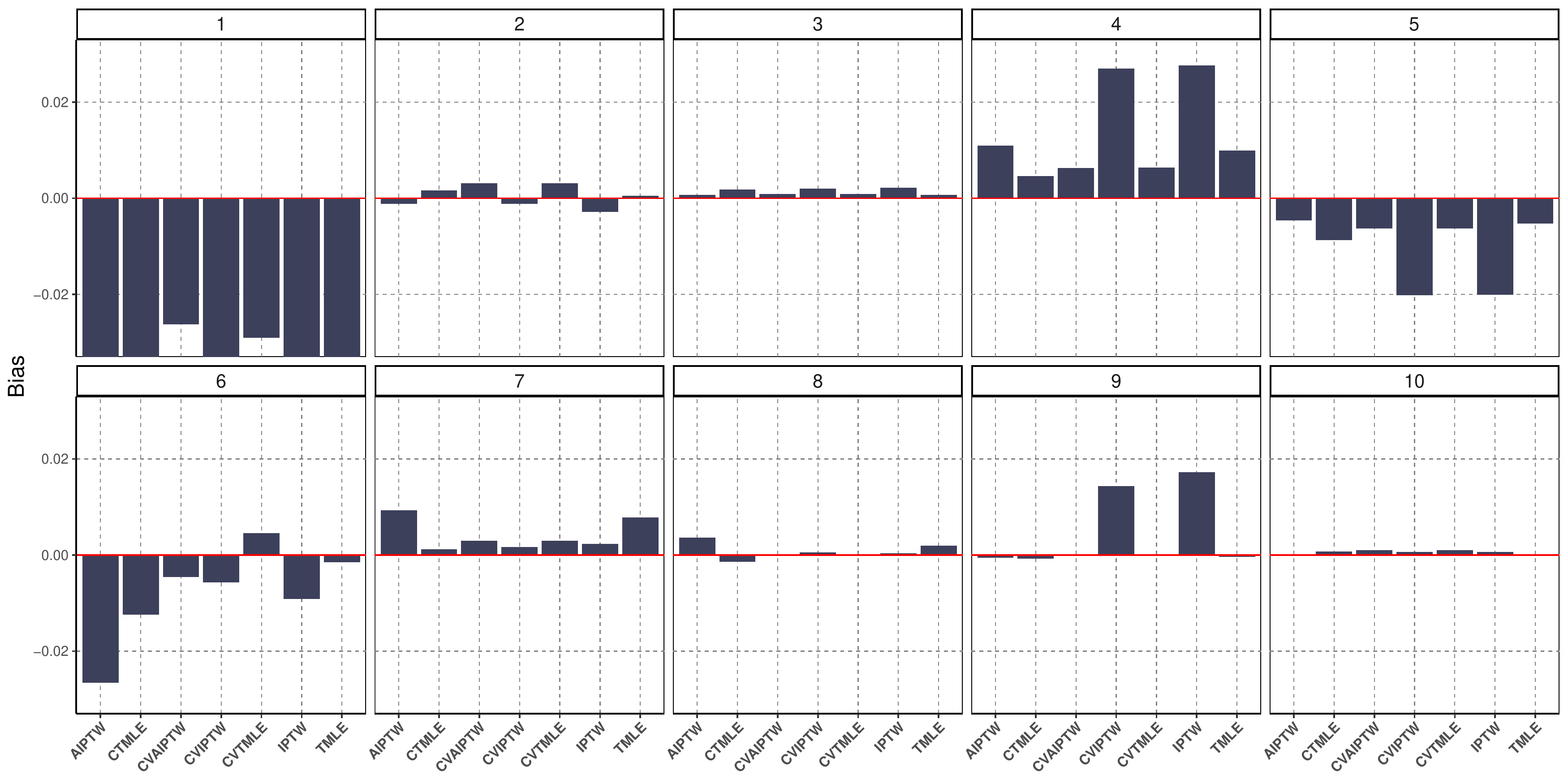}
\caption{\label{Figure 17} \text{Bias of estimators in ten studies}}
\end{figure}

\begin{figure}[!ht]
\centering\includegraphics[
  width=15cm,
  height=10cm,
  keepaspectratio
] {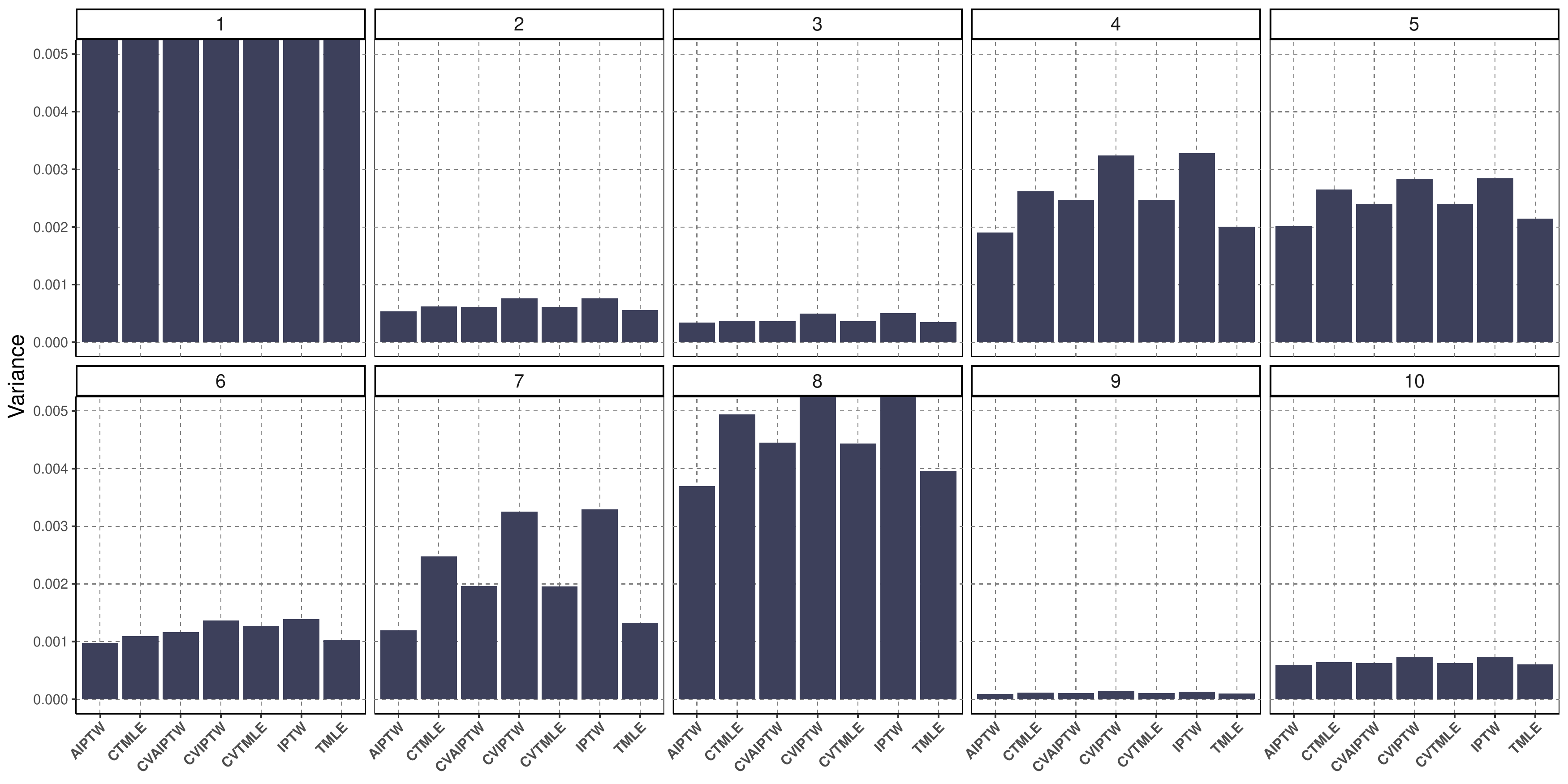}
\caption{\label{Figure 8} \text{Variance of estimators in ten studies}}
\end{figure}

\begin{figure}[!ht]
\centering\includegraphics[
  width=15cm,
  height=10cm,
  keepaspectratio
] {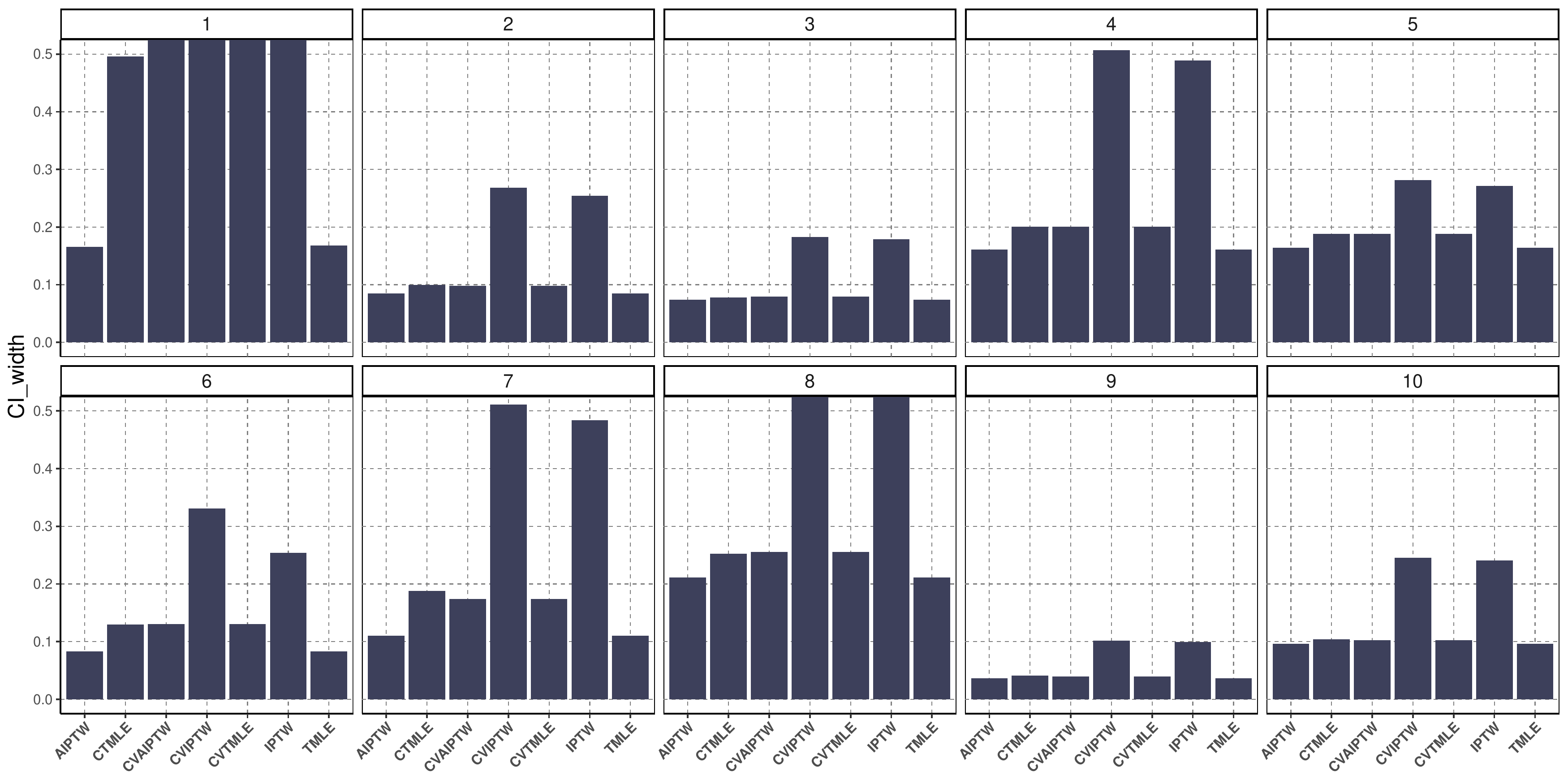}
\caption{\label{Figure 9} \text{Average width of confidence intervals in ten studies}}
\end{figure}

\begin{figure}[!ht]
\centering\includegraphics[
  width=15cm,
  height=10cm,
  keepaspectratio
] {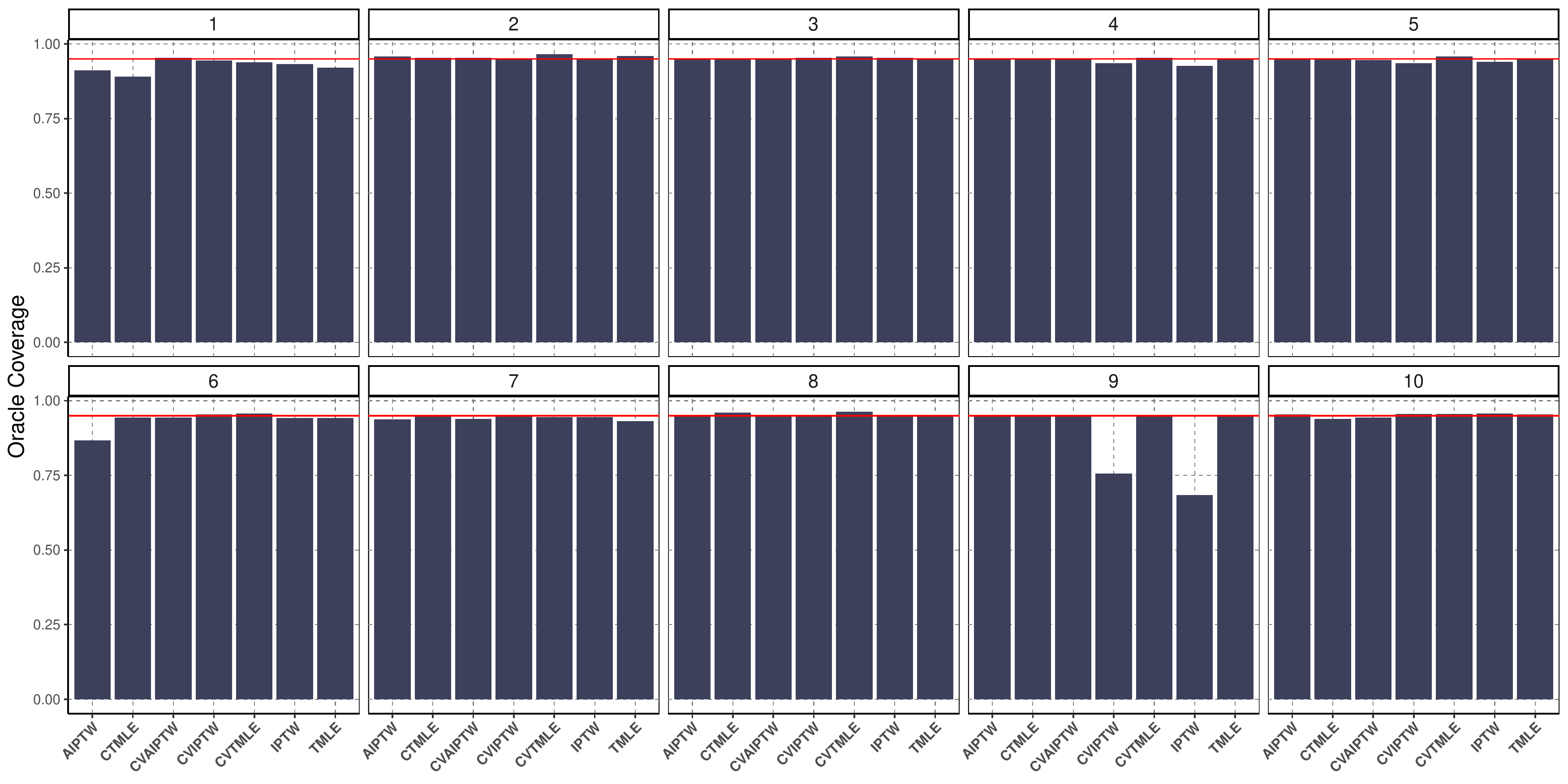}
\caption{\label{Figure 10} \text{Coverage of oracle confidence intervals in ten studies}}
\end{figure}

\section{.  Supplemental information on studies\label{app2}}%

\subsection{iLiNS-Zinc}
The iLiNS Zinc intervention was a placebo-controlled, cluster-randomized trial that enrolled children in Burkina Faso \cite{HessSLNS2015}. Data used in this analysis was collected between 2010 and 2012 from 3,265 children in rural southwestern Burkina Faso. The objective of this study was to compare the effect of providing small-quantity lipid-based nutrient supplements (SQ-LNS) with varying amounts of zinc, along with illness treatment, to standard care on zinc-related outcomes. The outcome measure used here was a height-to-age z-score for children. For the purposes of this analysis, the treatment arms were split into a control group (placebo) and a treatment group (zinc supplementation or SQ-LNS with varying amounts of zinc). 

\subsection{iLiNS-DOSE}
The iLiNS-DOSE intervention was a randomized, controlled, single-blind, parallel-group clinical trial that enrolled healthy infants between 5.5 and 6.5 months of age in Malawi \cite{MaletaP1LNA2015}. Data used in this analysis was collected between 2009 and 2011 for 1,931 children in rural Malawi. The objective of this trial was to identify the lowest growth-promoting daily dose of modified lipid-based nutrient supplements. The outcome measure used here was the children's height-to-age z-score. For the purposes of this analysis, treatment arms were split into a control group (no supplement during primary follow-up period with delayed supplementation from 18 to 30 months) and a treatment group consisting of multiple interventions (varying doses of milk-containing LNSs or milk-free LNSs between 6 and 18 months of age).

\subsection{JiVitA-3: Impact of antenatal multiple micronutrient supplementation on infant mortality}
The JiVitA-3 intervention was a cluster-randomized, double-masked trial that enrolled pregnant women from rural Bangladesh \cite{WestEMMMD2014}; data used in this analysis was collected between 2008 and 2012. The objective of this trial was to compare the effects of daily maternal multiple micronutrient versus iron and folic acid supplementation on 6-month infant mortality and adverse birth outcomes. The outcome measure used here was a height-to-age z-score for infants. Treatment arms were split into a control group (folic acid and iron supplementation only) and a treatment group (supplement containing 15 essential vitamins and minerals).

\subsection{JiVitA-4: Effect of fortified complementary food supplementation on child growth in rural Bangladesh}
The JiViTa-4 intervention was an unblinded, cluster-randomized, controlled trial that enrolled children at 6 months of age and provided daily supplements for a year \cite{ChristianEFCFD2015}. Data used in this analysis was collected between 2012 and 2014 for 5,443 children in rural Bangladesh. The objective of this trial was to compare the effects of three specially formulated complementary food supplements and an international product, Plumpy'doz, with a control (no food supplement) on children's growth outcomes. The outcome measure used here was a height-to-age z-score for children. For the purposes of this analysis, treatment arms were split into a control (nutrition counseling only) and a treatment group (food supplements with nutrition counseling). 

\subsection{WASH Benefits Bangladesh }
The WASH Benefits Bangladesh intervention was a cluster randomized trial that enrolled pregnant women from villages in rural Bangladesh \cite{LubyEWQSM2018}; data used in this analysis was collected between 2012 and 2015 for 4,863 children. The outcome measure used here was a height-to-age z-score for children. For the purposes of this analysis, treatment arms were split into a control group (data collection only) and a treatment group. This treatment group consists of multiple interventions: chlorinated drinking water; upgraded sanitation; promotion of handwashing with soap; combination of water, sanitation, and handwashing; child nutrition counseling plus lipid-based supplements; and a combination of water, sanitation, handwashing, and nutrition.

\subsection{WASH Benefits Kenya}
The WASH Benefits Kenya intervention was a cluster randomized trial that enrolled pregnant women from western Kenya \cite{StewartEWQSA2018}; data used in this analysis was collected between 2013 and 2015 for 7,399 children. The outcome measure used here was a height-to-age z-score for children. For the purposes of this analysis, treatment arms were split into a control group (no visits) and a treatment group. This treatment group consists of multiple interventions: chlorinated drinking water; improved sanitation; handwashing with soap; combined water, sanitation, and handwashing; nutrition counseling plus lipid-based supplements; and combined water, sanitation, handwashing, and nutrition.

\subsection{SAS Food Supplementation}
The SAS Food Supplementation intervention was a randomized trial that enrolled four-month-old infants in an urban slum in Delhi, India \cite{BhandariFSEFJ2001}. Data for this analysis was collected between 1995 and 1996 from 418 children. The objective of this study was to measure the effect of feeding a micronutrient-fortified food supplement to infants on physical growth between 4 and 12 months of age. The measured outcome used in this analysis was a height-to-age z-score for children. For the purposes of this analysis, treatment arms were split into a control group (no counseling or food supplements) and a treatment group consisting of multiple interventions (milk-based cereal and nutritional counseling, nutritional counseling alone, or visitation alone).

\subsection{iLiNS-DYAD-M}
The iLiNS-DYAD-M intervention was a randomized trial that enrolled infants 6 to 18 months of age in rural Malawi \cite{AshornSMDPJ2015}. Data was collected between 2011 and 2015 from 1,205 children. The objective of this study was to test whether small-quantity lipid-based nutrient supplements (SQ-LNS) would promote child growth with an outcome measure of the children's height-to-age z-score at 18 months. Treatment arms were split into a control group (iron and folate supplementation during pregnancy) and a treatment group with multiple interventions (multiple micronutrient tablets or small-quantity lipid-based nutrient supplements (SQ-LNS) during pregnancy and first six months of lactation). Children in the SQ-LNS group received SQ-LNSs from 6 to 18 months of age, while children in the control group and multiple-micronutrient tablet group did not receive any supplementation.

\subsection{Tanzania Child 2}
The Trial of Zinc and Micronutrients in Tanzanian Children was a randomized 2 x 2 factorial, double-blind clinical trial \cite{LocksEZMSM2016}. Data used in this analysis was collected from 2,396 children born to HIV-negative mothers in Tanzania. The secondary objective of this study was to determine whether daily administration of zinc or multivitamins from 6 weeks of age for 18 months would improve child growth; the measured outcome used in this analysis was the children's height-to-age z-score. For the purposes of this analysis, treatment arms were split into a control group (placebo) and a treatment group consisting of multiple interventions (zinc only, multivitamins only, or zinc and multivitamins).

\subsection{Lungwena Child Nutrition Intervention}
The Lungwena Child Nutrition Intervention was a single-blind, parallel randomized trial that enrolled six-month-old healthy infants in the Lungwena area of the Mangochi District in rural Malawi \cite{ThakwalakwaPTCMS2009}. Data for this analysis was collected between 2008 and 2014 from 840 children. The objective of this study was to determine whether lipid-based nutrient supplementation as a complementary food from 6 months to 18 months of age led to lower incidence of severe stunting in infants. The measured outcome used in this analysis was a height-to-age z-score for children. For the purposes of this analysis, treatment arms were split into a control group (no extra food supplements) and a treatment group consisting of multiple interventions (milk-powder-containing LNS, soy-powder protein supplement, or maize-soy flour supplement).

\bibliography{main.bib}

\clearpage

\end{document}